\documentclass[prd,aps,showkeys,floats,onecolumn]{revtex4}
\usepackage{amsmath}
\usepackage{amssymb}

\begin{document}

\title[Probing the origin of inertia]{Probing the origin
of inertia behind spacetime deformation}

\author{G Ter-Kazarian}

\address{Byurakan Astrophysical Observatory,
Byurakan 378433, Aragatsotn District, Armenia, E-mail:
gago\_50@yahoo.com}

\begin{abstract}
To investigate the origin and nature of inertia, we introduce a new
concept of hypothetical 2D, so-called, {\it master-space} (MS),
subject to certain rules. The MS, embedded in the background
4D-spacetime, is an indispensable individual companion to the
particle of interest, without relation to every other particle. We
argue that a deformation/(distortion of local internal properties)
of MS is the origin of inertia. With this perspective in sight, we
construct the alternative {\it relativistic theory of inertia}
(RTI), which allows to compute the {\em relativistic inertial force}
acting on an arbitrary point-like observer due to its {\em absolute
acceleration}. We go beyond the hypothesis of locality with an
emphasis on distortion of MS, which allows to improve essentially
the standard metric and other relevant geometrical structures
related to the noninertial reference frame of an arbitrary
accelerated observer. We compute the inertial force exerted on the
photon in a gravitating system in the semi-Riemann space. Despite
the totally different and independent physical sources of
gravitation and inertia, this approach furnishes justification for
the introduction of the principle of equivalence. Consequently, we
relate the inertia effects to the more general post-Riemannian
geometry. We derive a general expression of the relativistic
inertial force exerted on the extended spinning body moving in the
Rieman-Cartan space.

\end{abstract}


\maketitle

\section{Introduction}
The current observations made in the Earth-Moon-Sun system
\cite{Gil}-\cite{HK}, or at galactic and cosmological scales
\cite{HLa}-\cite{Ni10}, probe more deeply the {\em weak} principle
of equivalence (PE), which establishes the independence of free-fall
trajectories of the internal composition and structure of bodies.
The inertia effects in fact are of vital interest also for the
phenomenological aspects of the problem of neutrino oscillations,
see e.g. \cite{SG}-\cite{CL}. All this has evoked the study of the
inertial effects in an accelerated and rotated frame of stationary
laboratories on Earth. As long as all relevant length scales in
feasible experiments are very small in relation to the huge
acceleration lengths of the tiny accelerations we usually
experience, the curvature of the wordline could be ignored and the
differences between observations by accelerated and comoving
inertial observers will also be very small. Therefore, it is a
long-established practice in physics to use the hypothesis of
locality for extension of the Lorentz invariance to accelerated
observers in Minkowski spacetime. This in effect replaces the
accelerated observer by a continuous infinity of hypothetical
momentarily comoving inertial observers along its wordline. In this
line, in 1990, Hehl and Ni proposed a framework to study the
relativistic inertial effects of a Dirac particle \cite{Hehl7}. Ever
since this question has become a major preoccupation of physicists,
see e.g. \cite{HL}-\cite{Hua}. Even this works out, it is still
reminds us of a puzzling underlying reality of inertia. Despite our
best efforts, all attempts to obtain a true knowledge of the
geometry related to the noninertial reference frames of an arbitrary
observer seem doomed, unless we find {\em a physical principle the
inertia might refer to}, and that a working alternative {\em
relativistic theory of inertia} (RTI) is formulated. Otherwise one
wanders in a darkness. However, it seems that the inertia displays
no any physical characteristics of gravitation, because there are
many controversies to question the validity of such a description
\cite{Syn}-\cite{Drev}. For example, the experiments
by~\cite{Coc}-\cite{Drev} tested the key question of anisotropy of
inertia stemming from the idea that the matter in our galaxy is not
distributed isotropically with respect to the earth, and hence if
the inertia is due to gravitational interactions, then the inertial
mass of a body will depend on the direction of its acceleration with
respect to the direction towards the center of our galaxy. If the
nuclear structure of $Li^{7}$ is treated as a single $P_{3/2}$
proton in a central nuclear potential, the variation $\Delta m$ of
mass with direction, if it exists, was found to satisfy
$\frac{\Delta m}{m}\leq 10^{-20}$. This proves that there is no
anisotropy of mass which is due to the effects of mass in our
galaxy. Moreover, unlike gravitation, a curvature arisen due to
acceleration of coordinate frame of interest, i.e. a "fictitious
gravitation" which can be globally removed by appropriate coordinate
transformations, relates to this coordinate system itself and does
not affect the other systems or matter fields all at once.

In a recent paper \cite{gago1}, we construct the two-step spacetime
deformation theory. Thereby, through a choice of the {\em
world-deformation tensor}, $\widetilde{\Omega}$, which we have at
our disposal, in general, we have a way to deform the spacetime
displayed a different post Riemannian spacetime structures as its
corollary. This allows to construct a consistent Einstein-Cartan
theory, with the {\em dynamical torsion}. It is the purpose of the
present paper to carry out some details of this program to probe the
origin and nature of the phenomenon of inertia. We ascribe the
inertia effects to the geometry itself but as having a nature other
than gravitation. We propose a new concept of hypothetical 2D,
so-called, {\em master-space} (MS), subject to certain rules. The
MS, embedded in the background 4D-space, is an indispensable
individual companion to the particle of interest, without relation
to the other matter. Namely, the particle has to live with
MS-companion as an intrinsic property. This together with the idea
that the inertia effects arise as a deformation/(distortion of local
internal properties) of MS, are the highlights of the RTI. This
allows to compute the {\em relativistic inertial force} acting on an
arbitrary observer due to its {\em absolute acceleration}. The
hypothesis of locality represents strict restrictions, because it
approximately replaces the distorted MS, by the flat MS.  We might
have to go beyond the hypothesis of locality with an emphasis on
distortion of MS. This we might expect will essentially improve the
standard metric, etc., related to the noninertial system of an
arbitrary observer in Minkowski spacetime . We will proceed
according to the following structure. In section 2, we explain our
view of what is the MS, and lay a foundation of the {\em
relativistic law of inertia} (RLI). In section 3, a general
deformation/distortion of MS is described. In section 4, starting
with the Minkowski background space $M_{4}$, we construct the RTI.
In section 5, in the framework of a distortion of MS, we compute the
improved metric and other relevant geometrical structures in
noninertial system of an arbitrary accelerating and rotating
observer in Minkowski spacetime. The case of semi-Riemann background
space $V_{4}$ is dealt with in section 6, where we give
justification for the introduction of the PE on the theoretical
basis. In section 7, we relate the RTI to more general
post-Riemannian geometry. The concluding remarks are presented in
section 8. We will be brief and often suppress the indices without
notice. Unless otherwise stated we take natural units, $h=c=1$.

\section{The hypothetical MS-companion}
The MS is the 2D Minkowski space, $M_{2}$:
\begin{equation}
\begin{array}{l}
M_{2}=R_{(+)}^{1}\oplus R_{(-)}^{1}.
\end{array}
\label{R1}
\end{equation}
The ingredient 1D-space $ R_{A}^{1}$ is spanned by the coordinates
$\eta^{A}$, where we use the "naked" capital Latin letters
$A,B,...=(\pm)$ to denote the world indices related to $M_{2}$. The
metric in $M_{2}$ is
\begin{equation}
\begin{array}{l}
\overline{g}=\overline{g}(\overline{e}_{A},\,\overline{e}_{B})\,\overline{\vartheta}^{A}\otimes
\overline{\vartheta}^{B},
\end{array}
\label{O1}
\end{equation}
where $\overline{\vartheta}^{A}=d\eta^{A}$ is the infinitesimal
displacement. The basis $\overline{e}_{A}$ at the point of interest
in $M_{2}$ consists of two real {\em null vectors}:
\begin{equation}
\begin{array}{l}
\overline{g}(\overline{e}_{A},\,\overline{e}_{B})\equiv<\overline{e}_{A},\,\overline{e}_{B}>={}^{*}o_{A
B}, \quad ({}^{*}o_{A B})=\left(
                                                       \begin{array}{cc}
                                                         0 & 1 \\
                                                         1 & 0 \\
                                                       \end{array}
                                                     \right)
.
\end{array}
\label{R6}
\end{equation}
The norm, $i\overline{d}\equiv d\hat{\eta}$, given in this basis
reads
$i\overline{d}=\overline{e}\overline{\vartheta}=\overline{e}_{A}\otimes\overline{\vartheta}^{A}$,
where $i\overline{d}$ is the tautological tensor field of type
(1,1),  $\overline{e}$ is a shorthand for the collection of the
2-tuplet $(\overline{e}_{(+)},\,\overline{e}_{(-)})$, and $
\overline{\vartheta}=\left(
                    \begin{array}{c}
                      \overline{\vartheta}{}^{(+)} \\
                      \overline{\vartheta}{}^{(-)} \\
                    \end{array}
                  \right).
$ We may equivalently use a temporal $q^{0}\in T^{1}$ and  a spatial
$q^{1}\in R^{1}$ variables $q^{r} (q^{0},\, q^{1}) (r=0,1),$ such
that
\begin{equation}
\begin{array}{l}
M_{2}=R^{1}\oplus T^{1}.
\end{array}
\label{R1}
\end{equation}
The norm, $i\overline{d}$,  now can be rewritten in terms of
displacement, $dq^{r}$, as
\begin{equation}
\begin{array}{l}
i\overline{d}=d\hat{q}= e_{0}\otimes dq^{0} + e_{1}\otimes dq^{1},
\label{RE1}
\end{array}
\end{equation}
where $e_{0}$ and $e_{1}$ are, respectively, the temporal and
spatial basis vectors:
\begin{equation}
\begin{array}{l}
\overline{g}(\overline{e}_{r},\,\overline{e}_{s})\equiv<\overline{e}_{r},\,\overline{e}_{s}>=o_{r
s}, \quad (o_{r s})=\left(
                                                       \begin{array}{cc}
                                                         1 & 0 \\
                                                         0 & -1 \\
                                                       \end{array}
                                                     \right).
\end{array}
\label{O2}
\end{equation}
The MS is assumed to be embedded in the background 4D space and the
motion of the individual particle is fully depends on the properties
of MS-companion. In fact, we assume the particle has to be moving
simultaneously in the  parallel {\em individual} $M_{2}$ space and
the ordinary 4D background space (either Minkowskian or Riemannian).
Let us, first, concentrate our attention on non-accelerated
observer, who for the position  of a free test particle in the flat
MS uses the inertial coordinate frame $S_{(2)}$, such that
\begin{equation}
\begin{array}{l}
v^{(\pm)}=\frac{d\eta^{(\pm)}}{dq^{0}}=\frac{1}{\sqrt{2}}(1\pm
v_{q}), \quad v_{q}=\frac{dq^{1}}{dq^{0}}=const.
\end{array}
\label{O3}
\end{equation}
Suppose the position of this particle in the 4D space $M_{4}$ is
specified by the coordinates $x^{l}(s)$ $ (l=0,1,2,3)$ with respect
to the axes of the inertial system $S_{(4)}.$  We may adjust the
systems $S_{(2)}$ and $S_{(4)}$ in such a way as the spatial axis
$\vec{e}_{q}\equiv e_{1}$ of $S_{(2)}$ lies along the velocity
$\vec{v}=\vec{e}_{v}|\vec{v}|$\phantom{a}
($\vec{e}_{q}\,||\,\vec{e}_{v}$), while the time axis
$\vec{e}_{0}\equiv e_{0}$ of $S_{2}$ be the time axis of  a comoving
inertial frame $S_{4}$, such that the time coordinates in the two
systems are taken the same, $q^{0}=x^{0}=t$, and that $v_{q}\geq
0$.\\
We now define the concepts of {\em absolute} and {\em relative}
states of the ingredient spaces $R_{A}^{1}$. The measure for these
states is the very magnitude of the velocity components $v^{A}$ of
the particle:

{\bf Definition:} $$\begin{array}{l}\mbox{\em The ingredient
space}\, R_{A}^{1} \,\mbox{\em of the individual MS-companion of the
particle}\\ \mbox{\em is said to be in}\, \left\{
          \begin{array}{c}
            \mbox{\em absolute (abs) state if}\quad  v^{A}=0, \\
            \mbox{\em relative (rel) state if} \quad v^{A}\neq 0. \\
          \end{array}
 \right.\end{array}
 $$
Therefore, the MS  can be realized either in the {\em semi-absolute}
state (rel, abs), or (abs, rel), or  in the {\em total relative }
state (rel, rel). It is remarkable that the {\em total-absolute}
state, (abs, abs), which is equivalent to the unobservable Newtonian
{\em absolute} two-dimensional spacetime, cannot be realized because
of the relation $v^{(+)}+v^{(-)}=\sqrt{2}$. An existence of the {\em
absolute} state of the $R_{A}^{1}$ is an immediate cause of the
light traveling in empty space $R^{1}$ along the $q$-axis with a
maximal velocity $v_{q}=c$ (we re-instate the factor (c)) in the
$(+)-$direction corresponding to the state $ (v^{(+)},\,0)
\Leftrightarrow $ (rel, abs), and in the $(-)-$direction
corresponding to the state $(0,\,v^{(-)}) \Leftrightarrow$ {(abs,
rel)}. The {\em absolute} state of $R_{A}^{1}$ manifests its {\em
absolute} character in the important for SR fact that the resulting
velocity of light in the empty space $R^{1}$ is the same in all
inertial frames $S_{(2)}$, $S'_{(2)}$, $S''_{(2)}$,..., i.e., in
empty space light propagates independently of the state of motion of
the source\,-\, if $v^{A}=0$ then $v^{A}{}'=v^{A}{}''=...=0.$ Since
the $v^{A}$ is the very key measure of a deviation from the {\em
absolute} state, we might expect that this has a substantial effect
in an alteration of the particle motion under the unbalanced force.
This observation allows us to lay forth the foundation of the
fundamental RLI as follows:

{\bf Conjecture} (RLI-Conjecture): {\em The non-zero local rate
$\varrho(\eta,m,f)$ of instantaneously change of a constant velocity
$v^{A}$ (both magnitude and direction) of a massive $(m)$ test
particle under the unbalanced net force $(f)$ is the immediate cause
of a deformation/(distortion of the local internal properties) of
MS:\phantom{a} $M_{2} \rightarrow \widetilde{\mathcal{M}}_{2}$}.\\
We can conclude therefrom that, unless MS is flat, a free particle
in 4D background space in motion of uniform speed in a straight line
tends to stay in this motion and a particle at rest tends to stay at
rest. In this way, the MS-companion, therefore, abundantly serves to
account for the state of motion of the particle in the 4D background
space. In going into practical details, the function
$\varrho(\eta,m,f)$ will be determined in section 4.

\section{The general spacetime deformation/distortion-complex}
Based on the work \cite{gago1}, we now extend the geometrical ideas
of the spacetime deformation as applied to the 2D deformation $M_{2}
\rightarrow \widetilde{\mathcal{M}}_{2}$. To start with, let $V_{2}$
be 2D semi-Riemann space, which has at each point a tangent space,
$\breve{T}_{\breve{\eta}}V_{2}$, spanned by the anholonomic
orthonormal frame field, $\breve{e}$, as a shorthand for the
collection of the 2-tuplet $(\breve{e}_{(+)},\,\breve{e}_{(-)})$,
where $\breve{e}_{a}=\breve{e}_{a}^{\phantom{a}\tilde{A}}\,
\breve{e}_{\tilde{A}}$, with the holonomic frame is given as
$\breve{e}_{\tilde{A}}= \breve{\partial}_{\tilde{A}}$. Here, we use
the first half of Latin alphabet $a, b, c, . . . = (\pm)$ to denote
the anholonomic indices related to the tangent space, and the
capital Latin letters with an over
${}^{\prime}\phantom{a}\tilde{}\phantom{a}{}^{\prime}$ -
$\tilde{A},\tilde{B},...=(\pm)$, to denote the holonomic world
indices related  either to the space  $V_{2}$ or
$\widetilde{\mathcal{M}}_{2}$. All magnitudes related to the space,
$V_{2}$, will be denoted with an over
${}^{\prime}\phantom{a}\breve{}\phantom{a}{}^{\prime}$. These then
define a dual vector, $\breve{\vartheta}$, of differential forms, $
\breve{\vartheta}=\left(
                    \begin{array}{c}
                      \breve{\vartheta}{}^{(+)} \\
                      \breve{\vartheta}{}^{(-)} \\
                    \end{array}
                  \right),
$ as a shorthand for the collection of the $\breve{\vartheta}{}^{b}
=\breve{e}{}^{b}_{\phantom{a}
\tilde{A}}\,\breve{\vartheta}{}^{\tilde{A}}$, whose values at every
point form the dual basis, such that $\breve{e}_{a}\,\rfloor\,
\breve{\vartheta}{}^{b}=\delta^{b}_{a}$, where $\rfloor$ denoting
the interior product. Namely, this is a $C^{\infty}$-bilinear map
$\rfloor:\Omega^{1}\rightarrow \Omega^{0}$ with $\Omega^{p}$ denotes
the $C^{\infty}$-modulo of differential p-forms on $V_{4}$. In
components
$\breve{e}_{a}^{\phantom{a}\tilde{A}}\,\breve{e}{}^{b}_{\phantom{a}\tilde{A}}=\delta^{b}_{a}$.
On the manifold, $V_{2}$, the tautological tensor field,
$i\breve{d}$, of type (1,1) can be defined which assigns to each
tangent space the identity linear transformation. Thus for any point
$\breve{\eta}\in V_{2}$, and any vector $\breve{\xi}\in
\breve{T}_{\breve{\eta}}V_{2}$, one has
$i\breve{d}(\breve{\xi})=\breve{\xi}$. In terms of the frame field,
the $\breve{\vartheta}{}^{a}$ give the expression for $i\breve{d}$
as
$i\breve{d}=\breve{e}\breve{\vartheta}=\breve{e}_{(+)}\otimes\breve{\vartheta}{}^{(+)}+
\breve{e}_{(-)}\otimes\breve{\vartheta}{}^{(-)}$, in the sense that
both sides yield $\breve{\xi}$ when applied to any tangent vector
$\breve{\xi}$ in the domain of definition of the frame field. We may
consider general transformations of the linear group, $GL(2, R)$,
taking any base into any other set of four linearly independent
fields. The notation, $\{\breve{e}_{a},\,\breve{\vartheta}{}^{b}\}$,
will be used below for general linear frames. The holonomic metric
can be defined in the semi-Riemann space, $V_{2}$, as
\begin{equation}
\begin{array}{l}
\breve{g}=\breve{g}_{\tilde{A}\tilde{B}}\,\breve{\vartheta}{}^{\tilde{A}}\otimes\breve{\vartheta}{}^{\tilde{B}}=
\breve{g}(\breve{e}_{\tilde{A}}, \,\breve{e}_{\tilde{B}})\,
\breve{\vartheta}{}^{\tilde{A}}\otimes\breve{\vartheta}{}^{\tilde{B}},
\end{array}
\label{R24}
\end{equation}
with components,
$\breve{g}_{\tilde{A}\tilde{B}}=\breve{g}(\breve{e}_{\tilde{A}},
\breve{e}_{\tilde{B}})$ in the dual holonomic base
$\{\breve{\vartheta}{}^{\tilde{A}}\}$. The anholonomic orthonormal
frame field, $\breve{e}$, relates $\breve{g}$ to the tangent space
metric, ${}^{*}o_{ab}$, by $ {}^{*}o_{ab} = \breve{g}(\breve{e}_{a},
\,\breve{e}_{b})=
\breve{g}_{\tilde{A}\tilde{B}}\,\breve{e}_{a}^{\phantom{a}\tilde{A}}\,\breve{e}_{b}^{\phantom{a}\tilde{B}}
$, which  has the converse
$\breve{g}_{\tilde{A}\tilde{B}}={}^{*}o_{ab}\,\breve{e}{}^{a}_{\phantom{a}\tilde{A}}\,\breve{e}{}^{b}_{\phantom{a}\tilde{B}}$
because
$\breve{e}_{a}^{\phantom{a}\tilde{A}}\,\breve{e}{}^{a}_{\phantom{a}\tilde{B}}=\delta^{\tilde{A}}_{\tilde{B}}$.
With this provision, we build up a general {\em distortion-complex}
(DC), yielding a distortion of the flat space $M_{2}$, and show how
it restores the {\em world-deformation tensor} $\widetilde{\Omega}$,
which still has to be put in \cite{gago1} by hand. The DC-members
are the invertible distortion matrix $D$, the tensor $Y$ and the
{\em flat-deformation tensor} $\Omega$. Symbolically,
$$DC\,\sim\,(\breve{D},\,\breve{Y},\,\Omega)\rightarrow \widetilde{\Omega}.
$$
The following two steps went into the principle foundation of a {\em
distortion of local
internal properties of MS}. \\
1) We assume that the linear frame
$(\overline{e}_{A};\,\overline{\vartheta}{}^{A})$, at given point
($p\in M_{2}$), is undergone the {\em distortion} transformations,
conducted by  $(\breve{D},\,\breve{Y})$ and $(D,\, Y)$,
respectively, relating to $V_{2}$ and $\widetilde{\mathcal{M}}_{2}$,
recast in the form
\begin{equation}
\begin{array}{l}
\breve{e}_{\tilde{A}}=\breve{D}{}_{\tilde{A}}^{B}\,\bar{e}_{B},\quad
\breve{\vartheta}{}^{\tilde{A}}=\breve{Y}{}^{\tilde{A}}_{B}\,\bar{\vartheta}{}^{B},\quad
e_{\tilde{A}}=D_{\tilde{A}}^{B}\,\bar{e}_{B},\quad
\vartheta^{\tilde{A}}=Y^{\tilde{A}}_{B}\,\bar{\vartheta}{}^{B}.
\end{array}
\label{D1}
\end{equation}
2) We write the norm, $d\widetilde{\hat\eta}\equiv id$, of the
infinitesimal displacement, $d\widetilde{\eta}{}^{\tilde{A}}$, on
the general smooth differential 2D-manifold,
$\widetilde{\mathcal{M}}_{2}$, in terms of the spacetime structures
of $V_{2}$:
\begin{equation}
\begin{array}{l}
id=e\,\vartheta=\widetilde{\Omega}{}_{\tilde{A}}^{\phantom{a}\tilde{B}}\,\breve{e}_{\tilde{B}}\otimes
\breve{\vartheta}{}^{\tilde{A}}=\Omega_{b}^{\phantom{a}a}\,\breve{e}_{a}\otimes\breve{\vartheta}{}^{b}=
e_{\tilde{C}}\otimes\vartheta^{\tilde{C}}=e_{a}\otimes\vartheta^{a}=
\Omega^{\phantom{a}B}_{A}\,\bar{e}_{B}\otimes\bar{\vartheta}{}^{A}\,\in\,\widetilde{\mathcal{M}}_{2},
\end{array}
\label{D3}
\end{equation}
where $e=\{e_{a}=e_{a}^{\phantom{a}\tilde{C}}\,e_{\tilde{C}}\}$ is
the frame field and
$\vartheta=\{\vartheta^{a}=e^{a}_{\phantom{a}\tilde{C}}\,\vartheta^{\tilde{C}}\}$
is the coframe field defined on $\widetilde{\mathcal{M}}_{2}$, such
that $e_{a}\,\rfloor\, \vartheta^{b}=\delta^{b}_{a}$. The
deformation tensors
$\widetilde{\Omega}{}^{\phantom{a}\tilde{B}}_{\tilde{A}}=
\pi^{\phantom{a}\tilde{C}}_{\tilde{A}}\,\pi_{\tilde{C}}^{\phantom{a}\tilde{B}}$,
and $\Omega^{\phantom{a}B}_{A}$ imply
\begin{equation}
\begin{array}{l}
\widetilde{\Omega}{}^{\phantom{a}\tilde{B}}_{\tilde{A}}=
\breve{D}{}_{\tilde{A}}^{C}\,\Omega^{\phantom{a}D}_{C}\,\breve{Y}{}^{\tilde{B}}_{D},\quad
\Omega^{\phantom{a}B}_{A}=Y^{\tilde{C}}_{A}\,D{}_{\tilde{C}}^{B}\,,
\end{array}
\label{D4}
\end{equation}
provided
\begin{equation}
\begin{array}{l}
D_{\tilde{C}}^{A}=\pi_{\tilde{C}}^{\phantom{a}\tilde{B}}\,\breve{D}{}_{\tilde{B}}^{A},\quad
Y^{\tilde{C}}_{B}=\pi_{\phantom{a}\tilde{A}}^{\tilde{C}}\,\breve{Y}{}^{\tilde{A}}_{B},
\end{array}
\label{D2}
\end{equation}
such that
\begin{equation}
\begin{array}{l}
e_{\tilde{C}}=\pi_{\tilde{C}}^{\phantom{a}\tilde{B}}\,\breve{e}_{\tilde{B}}\equiv
\widetilde{\partial}_{\tilde{C}},\quad
\vartheta^{\tilde{C}}=\pi^{\tilde{C}}_{\phantom{a}\tilde{A}}\,\breve{\vartheta}{}^{\tilde{V}}\equiv
d\,\widetilde{\eta}^{\tilde{C}},\quad
\widetilde{\eta}^{\tilde{C}}\,\in\, {\cal U}\in
\widetilde{\mathcal{M}}_{2}.
\end{array}
\label{R29}
\end{equation}
Hence the anholonomic deformation tensor,
$\Omega^{\phantom{a}a}_{b}=\pi^{\phantom{a}a}_{c}\,\pi^{\phantom{a}c}_{b}=
\widetilde{\Omega}{}^{\phantom{a}\tilde{B}}_{\tilde{A}}\,\breve{e}{}^{a}_{\phantom{a}\tilde{B}}\,\breve{e}_{b}^{\phantom{a}\tilde{A}}$,
yields local tetrad deformations
\begin{equation}
\begin{array}{l}
e_{c}=\pi^{\phantom{a}a}_{c}\,\breve{e}_{a},\quad
\vartheta^{c}=\pi_{\phantom{a}b}^{c}\,\breve{\vartheta}{}^{b},\quad
e\,\vartheta=e_{a}\otimes\vartheta^{a}=\Omega^{a}_{\phantom{a}b}\,\breve{e}_{a}\otimes\breve{\vartheta}{}^{b}.
\end{array}
\label{R30}
\end{equation}
The matrices,
$\pi(\widetilde{\eta}):\phantom{a}=(\pi^{\phantom{a}a}_{b})(\widetilde{\eta})$,
are referred to as the {\em first deformation matrices}, and the
matrices $
\gamma_{cd}(\widetilde{\eta})={}^{*}o_{ab}\,\pi^{\phantom{a}a}_{c}(\widetilde{\eta})\,\pi_{d}^{\phantom{a}b}(\widetilde{\eta}),
$ \,-\, {\em second deformation matrices}. The matrices, $
\pi_{\phantom{a}c}^{a}(\widetilde{\eta})\,\in\, GL(2, R)\,\forall\,
\widetilde{\eta}, $ in general, give rise to right cosets of the
Lorentz group, i.e. they are the elements of the quotient group
$GL(2, R)/SO(1,1)$, because the Lorentz matrices, $\Lambda^{r}_{s}$,
 $(r,s=1,0)$ leave the Minkowski metric invariant. A right-multiplication of
$\pi(\widetilde{\eta})$ by a Lorentz matrix gives an other
deformation matrix. So, all the fundamental geometrical structures
on deformed/distorted MS in fact - the metric as much as the
coframes and connections - acquire a {\em deformation/distortion}
induced theoretical interpretation. If we deform the tetrad
according to (\ref{R30}), in general, we have two choices to recast
metric as follows: either writing the deformation of the metric in
the space of tetrads or deforming the tetrad field:
\begin{equation}
\begin{array}{l}
g={}^{*}o_{ab}\,\pi_{\phantom{a}c}^{a}\pi_{\,\,\,d}^{b}\breve{\vartheta}{}^{c}\otimes
\breve{\vartheta}{}^{d}= \gamma_{cd}\,\breve{\vartheta}{}^{c}\otimes
\breve{\vartheta}{}^{d}={}^{*}o_{ab}\,\vartheta^{a}\otimes
\vartheta^{b}.
\end{array}
\label{R34}
\end{equation}
In the first case, the contribution of the Christoffel symbols,
constructed by the metric $\gamma_{ab}$, reads
\begin{equation}
\begin{array}{l}
\Gamma_{\phantom{a}bc}^{a}=\frac{1}{2}\,\left(\breve{C}{}_{\phantom{a}bc}^{a}-\gamma^{aa'}\,\gamma_{bb'}\,\breve{C}{}_{\phantom{a}a'c}^{b'}-
    \gamma^{aa'}\,\gamma_{cc'}\,\breve{C}{}_{\phantom{a}a'b }^{c'} \right)
+ \frac{1}{2}\,\gamma^{aa'}\,\left(\breve{e}_{c} \,\rfloor\,
d\,\gamma_{ba'}- \breve{e}_{b}\,\rfloor\, d\,\gamma_{ca'}
     - \breve{e}_{a'} \,\rfloor \,d\,\gamma_{bc}\right).
\end{array}
\label{R35}
\end{equation}
The deformed metric can be split as \cite{gago1}:
\begin{equation}
\begin{array}{l}
g_{\tilde{A}\tilde{B}}(\pi)=\Upsilon^{2}(\pi)\,\breve{g}_{\tilde{A}\tilde{B}}+\gamma_{\tilde{A}\tilde{B}}(\pi),
\end{array}
\label{R38}
\end{equation}
where $\Upsilon(\pi)=\pi^{a}_{a}$, and
\begin{equation}
\begin{array}{l}
\gamma_{\tilde{A}\tilde{B}}(\pi)=[\gamma_{ab}-\Upsilon^{2}(\pi)\,{}^{*}o_{ab}]\,
\breve{e}{}^{a}_{\phantom{a}\tilde{A}}\,\breve{e}{}^{b}_{\phantom{a}\tilde{B}}.
\end{array}
\label{Rg}
\end{equation}
In the second case, we may write the commutation table for the
anholonomic frame, $\{e_{a}\}$,
\begin{equation}
\begin{array}{l}
[e_{a},\, e_{b}]=- \frac{1}{2}\,C_{\phantom{a}ab}^{c}\,e_{c},
\end{array}
\label{R26A3}
\end{equation}
and define the anholonomy objects
\begin{equation}
\begin{array}{l}
C_{\phantom{a}bc}^{a}= \pi^{a}_{\phantom{a}e}\,
{\pi^{-1}}_{\phantom{d}b}^{d} \,{\pi^{-1}}_{\phantom{f}c}^{f}\,
\breve{C}{}_{\,df}^{e}+2\,\pi^{a}_{\phantom{a}f}\,\breve{e}_{g}^{\phantom{a}\tilde{A}}
\,\left({\pi^{-1}}_{\phantom{g}\,[b}^{g}\partial{}_{\tilde{A}}\,{\pi^{-1}}_{\phantom{f}c]}^{f}\right).
\end{array}
\label{R45}
\end{equation}
Taking into account (\ref{D3}), the metric (\ref{R34})  can be
alternatively written in a general form of the spacetime or frame
objects:
\begin{equation}
\begin{array}{l}
 g= g_{\tilde{A}\tilde{B}}\,\vartheta^{\tilde{A}}\otimes
\vartheta^{\tilde{B}}=
\left(\widetilde{\Omega}{}_{\tilde{A}}^{\phantom{a}\tilde{B}}\,
\widetilde{\Omega}{}_{\tilde{C}}^{\phantom{a}\tilde{D}}\right)\,
\breve{g}_{\tilde{B}\tilde{D}}
\,\breve{\vartheta}{}^{\tilde{A}}\otimes\breve{\vartheta}{}^{\tilde{C}}=
{}^{*}o_{ab}\,\vartheta^{a}\otimes \vartheta^{b}=
(\Omega_{a}^{\phantom{a}c}\,\Omega_{b}^{\phantom{a}d})
{}^{*}o_{cd}\,\breve{\vartheta}{}^{a}\otimes
\,\breve{\vartheta}{}^{b}=
\gamma_{cd}\,\breve{\vartheta}{}^{c}\otimes
\,\breve{\vartheta}{}^{d}= \\\left(\Omega^{\phantom{a}
C}_{A}\,\Omega^{\phantom{a}
D}_{B}\right){}^{*}o_{CD}\,\bar{\vartheta}{}^{A}\otimes\bar{\vartheta}{}^{B}.
\end{array}
\label{D5}
\end{equation}
A significantly more rigorous formulation of the spacetime
deformation technique as we have presented it may be found in
\cite{gago1}.

\section{Model building in the 4D background Minkowski spacetime}
In this section, we construct the RTI in particular case when the
relativistic test particle accelerated in the Minkowski 4D
background flat space, $M_{4}$, under an unbalanced net force other
than gravitational. Here and henceforth we simplify DC for our use
by imposing the constraints
\begin{equation}
\begin{array}{l}
D_{\tilde{C}}^{A}=\breve{D} {}_{\tilde{B}}^{A},\quad
\breve{Y}{}^{\tilde{A}}_{B}=\breve{D} {}^{\tilde{A}}_{B},
\end{array}
\label{D6}
\end{equation}
and, therefore,
$$DC\,\sim\,(D,\,\Omega)\rightarrow \widetilde{\Omega}.$$
The (\ref{D4}), by virtue of (\ref{D3}) and (\ref{D6}), gives
\begin{equation}
\begin{array}{l}
\widetilde{\Omega}{}^{\phantom{a}\tilde{B}}_{\tilde{A}}=
\breve{D}{}_{\tilde{A}}^{C}\,\Omega^{\phantom{a}D}_{C}\,\breve{D}{}^{\tilde{B}}_{D}=
\pi^{\phantom{a}\tilde{B}}_{\tilde{A}},\quad
Y^{\tilde{C}}_{B}=\widetilde{\Omega}{}_{\tilde{A}}^{\phantom{a}\tilde{C}}\,\breve{D}{}^{\tilde{A}}_{B},
\end{array}
\label{D7}
\end{equation}
where the deformation tensor,
$\widetilde{\Omega}{}^{\phantom{a}\tilde{B}}_{\tilde{A}}$,  yields
the partial holonomic frame transformations
\begin{equation}
\begin{array}{l}
e_{\tilde{C}}=\breve{e}_{\tilde{C}},\quad
\vartheta^{\tilde{C}}=\widetilde{\Omega}{}^{\tilde{C}}_{\phantom{a}\tilde{A}}\,\breve{\vartheta}{}^{\tilde{V}},
\end{array}
\label{D8}
\end{equation}
or, respectively, the $\Omega^{\phantom{a}a}_{b}$ yields the partial
local tetrad deformations
\begin{equation}
\begin{array}{l}
e_{c}=\breve{e}_{c},\quad
\vartheta^{c}=\Omega_{\phantom{a}b}^{c}\,\breve{\vartheta}{}^{b},\quad
e\,\vartheta=e_{a}\otimes\vartheta^{a}=\Omega^{a}_{\phantom{a}b}\,\breve{e}_{a}\otimes\breve{\vartheta}{}^{b}.
\end{array}
\label{D9}
\end{equation}
Hence, (\ref{D3}) defines a diffeomorphism
$\widetilde{\eta}{}^{\tilde{A}}(\eta):\phantom{a}M_{2}\rightarrow
\widetilde{\mathcal{M}}_{2}$:
\begin{equation}
\begin{array}{l}
e_{\tilde{A}}\,Y^{\tilde{A}}_{B}=
\Omega^{\phantom{a}A}_{B}\,\overline{e}_{A}\,,
\end{array}
\label{R3}
\end{equation}
where $Y^{\tilde{A}}_{B}=\partial
\widetilde{\eta}^{\tilde{A}}/\partial \eta^{B}$. The conditions of
integrability, $
\partial_{A}\,Y^{\tilde{C}}_{B}=\partial_{B}\,Y^{\tilde{C}}_{A},
$ and non-degeneracy, $det|Y^{\tilde{A}}_{B}| \neq 0$, immediately
define a general form of the {\em flat-deformation tensor} $
\Omega^{\phantom{a}A}_{B}:\phantom{a}=
{D}_{\tilde{C}}^{A}\,\partial_{B} \Theta^{\tilde{C}}, $ where
$\Theta^{\tilde{C}}$ is an arbitrary holonomic function. To make the
remainder of our discussion a bit more concrete, it proves necessary
to provide, further, a constitutive ansatz of simple, yet tentative,
linear {\em distortion transformations}, which, according to
RLI-Conjecture, can be written in terms of {\em local rate}
$\varrho(\eta,m,f)$ of instantaneously change of the measure $v^{A}$
of massive $(m)$ test particle under the unbalanced net force $(f)$:
\begin{equation}
\begin{array}{l}
e_{\tilde{(+)}}(\varrho)=D^{\phantom{a} B}_{\tilde{(+)}}(\varrho)\,
\overline{e}_{B}=\overline{e}_{(+)}-\varrho(\eta,m,f)\,
v^{(-)}\,\overline{e}_{(-)},\quad
e_{\tilde{(-)}}(\varrho)=D^{\phantom{a} B}_{\tilde{(-)}}(\varrho)\,
\overline{e}_{B}=\overline{e}_{(-)}+\varrho(\eta,m,f)\,
v^{(+)}\,\overline{e}_{(+)}.
\end{array}
\label{R35}
\end{equation}
Clearly, these transformations imply a violation of the relation
(\ref{R6})\, ($e_{\tilde{A}}^{2}(\varrho)\neq 0$) for the null
vectors $\overline{e}_{A}$.  The ~(\ref{D3}), for dual vectors of
differential forms  $\quad \vartheta= \left(
  \begin{array}{c}
    \vartheta^{(\tilde{+})} \\
    \vartheta^{(\tilde{-})} \\
  \end{array}
\right) $ and $\quad \overline{\vartheta}= \left(
  \begin{array}{c}
    \overline{\vartheta}{}^{(+)} \\
    \overline{\vartheta}{}^{(-)} \\
  \end{array}
\right)$, gives
\begin{equation}
\vartheta=\left(
            \begin{array}{cc}
              \Omega^{\phantom{a}C}_{(+)}<e^{\tilde{(+)}},\,\overline{e}_{C}> \,\,&\,\, \Omega^{\phantom{a}C}_{(-)}<e^{\tilde{(+)}},\,\overline{e}_{C}> \\
              \Omega^{\phantom{a}C}_{(+)}<e^{\tilde{(-)}},\,\overline{e}_{C}> \,\,&\,\, \Omega^{\phantom{a}C}_{(-)}<e^{\tilde{(-)}},\,\overline{e}_{C}> \\
            \end{array}
          \right)
\,\overline{\vartheta}. \label{M1}
\end{equation}
We may parameterize the tensor $ \Omega^{\phantom{a}A}_{B}$ in terms
of the parameters $\tau_{1}$ and $\tau_{2}$ as
\begin{equation}
\begin{array}{l}
\Omega^{\phantom{a}(+)}_{(+)}=\Omega^{\phantom{a}(-)}_{(-)}=\tau_{1}(1+\tau_{2}\,\overline{\varrho}{}^{2}),\quad
\Omega^{\phantom{a}(-)}_{(+)}=-\tau_{1}(1-\tau_{2})\varrho
v^{(-)},\quad
\Omega^{\phantom{a}(+)}_{(-)}=\tau_{1}(1-\tau_{2})\varrho v^{(+)},
\end{array}
\label{M2}
\end{equation}
where  $\overline{\varrho}{}^{2}=v^{2}\varrho^{2},\quad
v^{2}=v^{(+)}v^{(-)}=1/2\gamma^{2}_{q}$ and $
\gamma_{q}=(1-v_{q}^{2})^{-1/2}$. Then, the relation (\ref{M1}) can
be recast in an alternative form
\begin{equation}
\begin{array}{l}
\vartheta= \tau_{1}\left(
  \begin{array}{cc}
    1 & -\tau_{2}\varrho\, v^{(+)}\\
    \tau_{2}\varrho\, v^{(-)} & 1 \\
  \end{array}
\right)\,\overline{\vartheta}. \label{RL5}
\end{array}
\end{equation}
Suppose a second observer, who makes measurements using a frame of
reference $\widetilde{S}_{(2)}$ which is held stationary in
deformed/distorted space $\widetilde{\mathcal{M}}_{2}$, uses for the
test particle the corresponding spacetime coordinates
$\widetilde{q}{}^{\tilde{r}}\left((\widetilde{q}{}^{\tilde{0}},\,
\widetilde{q}{}^{\tilde{1}})\equiv (\widetilde{t},\,\widetilde{q}
)\right)$. The (\ref{D3}) can be rewritten in terms of spacetime
variables as
\begin{equation}
\begin{array}{l}
id=e\,\vartheta\equiv d\widetilde{\hat{q}}=\widetilde{e}_{0}\otimes
d\widetilde{t} + \widetilde{e}_{q}\otimes d\widetilde{q},
\label{RE4}
\end{array}
\end{equation}
where $\widetilde{e}_{0}$ and $\widetilde{e}_{q}$ are, respectively,
the temporal and spatial basis vectors:
\begin{equation}
\begin{array}{l}
\widetilde{e}_{0}(\varrho)=
\frac{1}{\sqrt{2}}\left[e_{\tilde{(+)}}(\varrho)+e_{\tilde{(-)}}(\varrho)\right],\quad
\widetilde{e}_{q}(\varrho)=
\frac{1}{\sqrt{2}}\left[e_{\tilde{(+)}}(\varrho)-e_{\tilde{(-)}}(\varrho)\right].
\end{array}
\label{O4}
\end{equation}
The transformation equation for the coordinates, according to
(\ref{RL5}), becomes
\begin{equation}
\begin{array}{l}
\vartheta^{(\tilde{\pm})}=\tau_{1} \,(\overline{\vartheta}{}^{(\pm)}
\mp \tau_{2}\,\varrho
v^{(\pm)}\overline{\vartheta}{}^{(\mp)})=\tau_{1}\, (v^{(\pm)}\mp
\tau_{2}\,\varrho v^{2})dt,
\end{array}
\label{RL7}
\end{equation}
which gives the general transformation equations for spatial and
temporal coordinates as follows ($\vec{e}_{q}\equiv e_{1},\,q\equiv
q^{1}$):
\begin{equation}
\begin{array}{l}
d\widetilde{t}= \tau_{1}\,dt,\quad d\widetilde{q}=
\tau_{1}\left[\,dq(1+\frac{\tau_{2}\varrho v_{q}
}{\sqrt{2}})-\frac{\tau_{2}\varrho
}{\sqrt{2}}dt\right]=\tau_{1}\,(dq-\frac{\tau_{2}\varrho
}{\sqrt{2}\gamma^{2}_{q}}dt).
\end{array}
\label{RL8}
\end{equation}
Hence, the general metric (\ref{D5}) in
$\widetilde{\mathcal{M}}_{2}$ reads
\begin{equation}
\begin{array}{l}
 g\equiv
d\widetilde{s}{}^{2}_{q}=g_{\tilde{r}\tilde{s}}\,d\widetilde{q}{}^{\tilde{r}}\otimes
d\widetilde{q}{}^{\tilde{s}} =
\left[(\Omega^{\phantom{a}(+)}_{(+)})^{2}+\Omega^{\phantom{a}(+)}_{(-)}\Omega^{\phantom{a}(-)}_{(+)}\right]ds^{2}_{q}+
\Omega^{\phantom{a}(+)}_{(+)}\left(\Omega^{\phantom{a}(+)}_{(-)}+
\Omega^{\phantom{a}(-)}_{(+)}\right)(dt\otimes dt+dq\otimes
dq)-\\2\Omega^{\phantom{a}(+)}_{(+)}\left(\Omega^{\phantom{a}(+)}_{(-)}-\Omega^{\phantom{a}(-)}_{(+)}\right)\,dt\otimes
dq,
\end{array}
\label{Tau2}
\end{equation}
provided
\begin{equation}
\begin{array}{l}
 g_{\tilde{0}\tilde{0}}=(1+\frac{\varrho
v_{q}}{\sqrt{2}})^{2}-\frac{\varrho^{2}}{2},\quad
g_{\tilde{1}\tilde{1}}=-(1-\frac{\varrho
v_{q}}{\sqrt{2}})^{2}+\frac{\varrho^{2}}{2},\quad
g_{\tilde{1}\tilde{0}}=g_{\tilde{0}\tilde{1}}=-\sqrt{2}\varrho.
\label{RE494}
\end{array}
\end{equation}
The difference of the vector, $d\hat{q}\in M_{2}$ (\ref{RE1}), and
the vector, $d\widetilde{\hat{q}}\in
\widetilde{\mathcal{M}}_{2}$~(\ref{RE4}), can be interpreted by the
second observer as being due to the deformation/distortion of flat
space $M_{2}$. However, this difference with equal justice can be
interpreted by him as a definite criterion for the {\em absolute}
character of his own state of acceleration in $M_{2}$, rather than
to any absolute quality of a deformation/distortion of $M_{2}$. To
prove this assertion, note that the transformation equations
(\ref{RL8}) give a reasonable change at low  velocities $v_{q}\simeq
0$, as
\begin{equation}
\begin{array}{l}
d\widetilde{t}= \tau_{1}\,dt,\quad d\widetilde{q}\simeq
\tau_{1}\,(dq-\frac{\tau_{2}\varrho }{\sqrt{2}}dt),
\end{array}
\label{O5}
\end{equation}
thereby
\begin{equation}
\begin{array}{l}
\Omega^{\phantom{a}(+)}_{(+)}=\Omega^{\phantom{a}(-)}_{(-)}=\tau_{1}(1+\tau_{2}\overline{\varrho}{}^{2}),\quad
\Omega^{\phantom{a}(+)}_{(-)}=-\Omega^{\phantom{a}(-)}_{(+)}=\tau_{1}(1-\tau_{2})\overline{\varrho}.
\end{array}
\label{O9}
\end{equation}
The (\ref{O5}) becomes conventional transformation equations to
accelerated $(a\neq 0)$ axes if we assume
$d(\tau_{2}\varrho)/\sqrt{2}dt= a$ and $\tau_{1}(v_{q}\simeq 0)=1$.
In high velocity limit $v_{q}\simeq 1$,\, $\overline{\varrho} \simeq
0,\, (d\eta^{(-)}= v^{(-)}dt\simeq 0,\ v^{(+)}\simeq v\simeq
\sqrt{2}$), we have
\begin{equation}
\begin{array}{l}
\Omega^{\phantom{a}(+)}_{(+)}=\Omega^{\phantom{a}(-)}_{(-)}=\tau_{1},\quad
\Omega^{\phantom{a}(-)}_{(+)}=0,\quad
\Omega^{\phantom{a}(+)}_{(-)}=\tau_{1}(1-\tau_{2})\sqrt{2}\varrho,
\end{array}
\label{O10}
\end{equation}
so (\ref{RL8}) and (\ref{Tau2}), respectively, give
\begin{equation}
\begin{array}{l}
d\widetilde{t}= \tau_{1}\,dt\simeq \tau_{1}\,dq\simeq
d\widetilde{q},
\end{array}
\label{Tau4}
\end{equation}
and
\begin{equation}
\begin{array}{l}
d\widetilde{s}{}^{2}_{q}\simeq
\left[(1+\frac{\varrho}{\sqrt{2}})^{2}-\frac{\varrho^{2}}{2}\right]d\widetilde{t}\otimes
d\widetilde{t}+
\left[-(1-\frac{\varrho}{\sqrt{2}})^{2}+\frac{\varrho^{2}}{2}\right]d\widetilde{q}\otimes
d\widetilde{q} -2\sqrt{2}\varrho \,d\widetilde{t}\otimes
d\widetilde{q} \simeq \tau_{1}^{2}\,ds^{2}_{q}=0.
\end{array}
\label{Tau6}
\end{equation}
To this end, {\em the inertial effects become zero}. Let
$\vec{a}_{net}$ be a local net 3-acceleration of an arbitrary
observer with proper linear 3-acceleration $\vec{a}$ and proper
3-angular velocity $\vec{\omega}$ measured in the rest frame:
\begin{equation}
\begin{array}{l}
\vec{a}_{net}=\frac{d\vec{u}}{ds}=\vec{a}\,\wedge\,\vec{u}
+\vec{\omega}\times \vec{u},
\end{array}
\label{G0}
\end{equation}
where ${\bf u}$ is the 4-velocity. A magnitude of $\vec{a}_{net}$
can be computed as the simple invariant of the absolute value
$|\frac{d{\bf u}}{ds}|$ as measured in rest frame:
\begin{equation}
\begin{array}{l}
|{\bf a}|=|\frac{d{\bf u}}{ds}|=
\left(\frac{du^{l}}{ds},\,\frac{du_{l}}{ds}\right)^{1/2}.
\end{array}
\label{G1}
\end{equation}
Following \cite{Syn, MTW}, let us define an orthonormal frame
$e_{\hat{a}}$, carried by an accelerated observer, who moves with
proper linear 3-acceleration and $\vec{a}(s)$ and proper 3-rotation
$\vec{\omega}(s)$.  Let the zeroth leg of the frame $e_{\hat{0}}$ be
4-velocity ${\bf u}$ of the observer that is tangent to the
worldline at a given event $x^{l}(s)$ and we parameterize the
remaining  spatial triad frame vectors $e_{\hat{i}}$, orthogonal to
$e_{\hat{0}}$, also by $(s)$. The spatial triad $e_{\hat{i}}$
rotates with proper 3-rotation $\vec{\omega}(s)$. The 4-velocity
vector naturally undergoes Fermi-Walker transport along the curve C,
which guarantees that $e_{\hat{0}}(s)$ will always be tangent to C
determined by $x^{l} = x^{l}(s)$:
\begin{equation}
\begin{array}{l}
\frac{de_{\hat{a}}}{ds}=-\Omega\, e_{\hat{a}}
\end{array}
\label{G2}
\end{equation}
where the antisymmetric rotation tensor $\Omega$ splits into a
Fermi-Walker transport part  $\Omega_{FW}$ and a spatial rotation
part $\Omega_{SR}$:
\begin{equation}
\begin{array}{l}
\Omega^{lk}_{FW}=a^{l}u^{k}-a^{k}u^{l},\quad
\Omega^{lk}_{SR}=u_{m}\omega_{n}\,\varepsilon^{mnlk}.
\end{array}
\label{G3}
\end{equation}
The 4-vector of rotation $\omega^{l}$ is orthogonal to 4-velocity
$u^{l}$, therefore, in the rest frame it becomes
$\omega^{l}(0,\,\vec{\omega})$, and $\varepsilon^{mnlk}$ is the
Levi-Civita tensor with $\varepsilon^{0123}=-1$. The (\ref{O5})
immediately indicates that we may introduce the very concept of the
local {\em absolute acceleration} (in Newton's terminology) brought
about via the Fermi-Walker transported frames as
\begin{equation}
\begin{array}{l}
\vec{a}_{abs}\equiv
\vec{e}_{q}\frac{d(\tau_{2}\varrho)}{\sqrt{2}ds_{q}}=\vec{e}_{q}\,|\frac{de_{\hat{0}}}{ds}|=\vec{e}_{q}\,|{\bf
a}|,
\end{array}
\label{RL888}
\end{equation}
where we choose the system $S_{(2)}$ in such a way as the axis
$\vec{e}_{q}$ lies along the net 3-acceleration
($\vec{e}_{q}\,||\,\vec{e}_{a}),\quad
(\vec{e}_{a}=\vec{a}_{net}/|\vec{a}_{net}|$). Combining (\ref{RL7})
and (\ref{RL888}), we obtain the key relation between a so-called
{\em inertial} acceleration
\begin{equation}
\begin{array}{l}
\vec{a}_{in}=\vec{e}_{a}\,
\frac{d^{2}\widetilde{q}}{ds^{2}_{q}}=\vec{e}_{a}
\frac{1}{\sqrt{2}}(\frac{d^{2}\widetilde{\eta}{}^{(+)}}{ds_{q}^{2}}-
\frac{d^{2}\widetilde{\eta}{}^{(-)}}{ds_{q}^{2}}),
\end{array}
\label{RL88}
\end{equation}
and a local {\em absolute acceleration} as follows:
\begin{equation}
\begin{array}{l}
\gamma_{q}\,\vec{a}_{in}=-\vec{a}_{abs}.
\end{array}
\label{RL999}
\end{equation}
The (\ref{RL999}) provides a quantitative means for the {\em
inertial force} $\vec{f}_{(in)}$:
\begin{equation}
\begin{array}{l}
\vec{f}_{(in)}=m\vec{a}_{in}= -\frac{m\vec{a}_{abs}}{\gamma_{q}},
\end{array}
\label{G33}
\end{equation}
In case of absence of rotation, we may write the local {\em absolute
acceleration} (\ref{RL888}) in terms of the relativistic force
$f^{l}$ acting on a particle with coordinates
$x^{l}(s)$~(\cite{W72}):
\begin{equation}
\begin{array}{l}
f^{l}(f^{0},\,\vec{f})=m\frac{d^{2}x^{l}}{ds^{2}}=
\Lambda^{l}_{k}(\vec{v})F^{k}.
\end{array}
\label{F1}
\end{equation}
Here $F^{k}(0,\,\vec{F})$ is the force defined in the rest frame of
the test particle, $\Lambda^{l}_{k}(\vec{v})$ is the Lorentz
transformation matrix $(i,j=1,2,3)$:
\begin{equation}
\begin{array}{l}
\Lambda^{i}_{j}=\delta_{i
j}-(\gamma-1)\frac{v_{i}v_{j}}{|\vec{v}|^{2}},\quad
\Lambda^{0}_{i}=\gamma v_{i},
\end{array}
\label{F2}
\end{equation}
where $ \gamma=(1-\vec{v}^{2})^{-1/2}.$  So the local rate
$\varrho(m,\,f^{l})$ of change of the measure of difference from the
{\em absolute} state for massive ($m$) test particle under the
unbalanced net force $f^{l}(x^{0},\,x^{i})( f^{0},\,\vec{f})$ other
than gravitational at the instant $x^{0}$ when the acceleration
begins, can be determined as
\begin{equation}
\begin{array}{l}
\frac{1}{\sqrt{2}}\frac{d(\tau_{2}\varrho)}{ds_{q}}=|{\bf
a}|=\frac{1}{m}|f^{l}|=
\frac{1}{m}(f^{l}f_{l})^{1/2}=\frac{1}{m\gamma} |\vec{f}|.
\end{array}
\label{RLL9}
\end{equation}
The (\ref{RL888}), (\ref{RLL9}) and (\ref{G3}) give
\begin{equation}
\begin{array}{l}
\vec{f}_{(in)}=-\frac{1}{\gamma_{q}\gamma}[\vec{F}+(\gamma-1)
\frac{\vec{v}(\vec{v}\cdot\vec{F})}{|\vec{v}|^{2}}].
\end{array}
\label{RLL999}
\end{equation}
At low velocities $v_{q}\simeq |\vec{v}|\simeq 0,$  the
~(\ref{RLL999}) reduces to the conventional non-relativistic law of
inertia
\begin{equation}
\begin{array}{l}
\vec{f}_{(in)}=-m\vec{a}_{abs}=-\vec{F}.
\end{array}
\label{RLG7}
\end{equation}
At high velocities $v_{q}\simeq|\vec{v}|\simeq 1$, if
$(\vec{v}\cdot\vec{F})\neq 0,$  the inertial force (\ref{RLL999})
becomes
\begin{equation}
\begin{array}{l}
\vec{f}_{(in)}\simeq-\frac{1}{\gamma}\vec{e}_{v}(\vec{e}_{v}\cdot\vec{F}),
\end{array}
\label{RKK}
\end{equation}
and, in agreement with (\ref{Tau6}), it vanishes in the limit of the
photon $(|\vec{v}|= 1, \,m=0).$  Thus, it takes force to disturb an
inertia state, i.e. to make the {\em absolute acceleration}
($\vec{a}_{abs}\neq 0$). The {\em absolute acceleration} is due to
the  real deformation/distortion of the space $M_{2}$. The {\em
relative} ($d(\tau_{2}\varrho)/ds_{q}=0$) acceleration (in Newton's
terminology) (both magnitude and direction), to the contrary, has
nothing to do with the deformation/distortion of the space $M_{2}$
and, thus, it cannot produce an inertia effects.

\section{The accelerating and rotating observer in Minkowski spacetime}
The standard geometrical structures, related to the noninertial
coordinate frame of accelerating and rotating observer in Minkowski
spacetime, were computed on the base of the hypothesis of locality
\cite{Hehl7}-\cite{MB2}), which in effect replaces an accelerated
observer at each instant with a momentarily comoving inertial
observer along its wordline. This assumption represents strict
restrictions, because in other words, it approximately replaces a
noninertial frame of reference $\widetilde{S}_{(2)}$, which is held
stationary in the deformed/distorted space
$\widetilde{\mathcal{M}}_{2}\equiv V_{2}^{(\varrho)}\, (\varrho\neq
0)$, with a continuous infinity set of the inertial frames
$\{\widetilde{S}_{(2)},\,\widetilde{S}'_{(2)},\,\widetilde{S}_{(2)}'',...\}$
given in the flat $M_{2}\, (\varrho=0)$. Therefore, it appears
natural to go beyond the hypothesis of locality with an emphasis on
distortion of MS, which we might expect will essentially improve the
standard results. The notation will be slightly different from the
previous section. We denote the orthonormal frame $e_{\hat{a}}$
(\ref{G2}), carried by an accelerated observer, with the over '{\em
breve}'
 such that
\begin{equation}
\begin{array}{l}
\breve{e}_{\hat{a}}=\overline{e}{}_{\phantom{a}\hat{a}}^{\,
\mu}\,\overline{e}_{\mu}=
\breve{e}{}_{\phantom{a}\hat{a}}^{\,\mu}\,\breve{e}_{\mu},\quad
\breve{\vartheta}{}^{\hat{b}}=\overline{e}{}^{\phantom{a}\hat{b}}_{
\mu}\,\overline{\vartheta}{}^{\mu}=
\breve{e}{}^{\phantom{a}\hat{b}}_{\,\mu}\,\breve{\vartheta}^{\mu},
\end{array}
\label{W01}
\end{equation}
with $\overline{e}_{\mu}=\partial_{\mu}=\partial/\partial
x^{\mu},\quad
\breve{e}_{\mu}=\breve{\partial}_{\mu}=\partial/\partial
\breve{x}{}^{\mu},\quad \overline{\vartheta}{}^{\mu}=dx^{\mu},\quad
\breve{\vartheta}^{\mu}=d\breve{x}. $ Here, following \cite{MTW,
MB1}, we introduced a geodesic coordinate system
$\breve{x}{}^{\mu}$\,-\, "coordinates relative to the accelerated
observer" (laboratory coordinates), in the neighborhood of the
accelerated path. The coframe members
$\{\breve{\vartheta}{}^{\,\hat{b}}\}$ are the objects of dual
counterpart:\,
$\breve{e}_{\hat{a}}\,\rfloor\,\breve{\vartheta}{}^{\hat{b}}=\delta^{b}_{a}$.
We choose the zeroth leg of the frame, $\breve{e}_{\hat{0}}$, as
before, to be the unit vector ${\bf u}$ that is tangent to the
worldline at a given event $x^{\mu}(s)$, where $(s)$ is a proper
time measured along the accelerated path by the standard (static
inertial) observers in the underlying global inertial frame. The
condition of orthonormality for the frame field
$\overline{e}{}^{\,\mu}_{\phantom{a}\hat{a}}$ reads
$\eta_{\mu\nu}\,\overline{e}{}^{\,\mu}_{\phantom{a}\hat{a}}\,\overline{e}{}^{\,\nu}_{\phantom{a}\hat{b}}=
o_{\hat{a}\hat{b}}=diag(+---)$. The antisymmetric acceleration
tensor $\Phi_{ab}$ \cite{MB1}-\cite{MF} is given by
\begin{equation}
\begin{array}{l}
\Phi_{a}^{\phantom{a}b}:\,=\overline{e}{}^{\phantom{a}\hat{b}}_{
\mu}\,\frac{d\overline{e}{}_{\phantom{a}\hat{a}}^{\, \mu}}{ds},
\end{array}
\label{WW1}
\end{equation}
where according to (\ref{G2}) and (\ref{G3}), and in analogy with
the Faraday tensor, one can identify $\Phi_{ab}\rightarrow (-{\bf
a},\,{\bf \omega})$, with ${\bf a}(s)$ as the translational
acceleration $\Phi_{0i}=-a_{i}$, and ${\bf \omega}(s)$ as the
frequency of rotation of the local spatial frame with respect to a
nonrotating (Fermi- Walker transported) frame
$\Phi_{ij}=-\varepsilon_{ijk}\,\omega^{k}$. The hypothesis of
locality holds for huge proper acceleration lengths $|I|^{-1/2}\gg
1$ and $|I^{*}|^{-1/2}\gg 1$, where the scalar invariants are given
by \,$I=(1/2)\,\Phi_{ab}\,\Phi^{ab}=-\vec{a}^{2}+\vec{\omega}^{2}$
and $I^{*}=(1/4)\,\Phi_{ab}^{*}\,\Phi^{ab}=-\vec{a}\cdot
\vec{\omega}$ ($\Phi_{ab}^{*}=\varepsilon_{abcd}\,\Phi^{cd}$)
\cite{MB1, MB2}. Suppose the displacement vector $z^{\mu}(s)$
represents the position of the accelerated observer. According to
the hypothesis of locality, at any time $(s)$ along the accelerated
worldline the hypersurface orthogonal to the worldline is Euclidean
space and we usually describe some event on this hypersurface
("local coordinate system") at $x^{\mu}$ to be at
$\breve{x}{}^{\mu}$, where $x^{\mu}$ and $\breve{x}^{\mu}$ are
connected via $\breve{x}{}^{\,0}= s$ and
\begin{equation}
\begin{array}{l}
x^{\mu}=z^{\mu}(s)+\breve{x}{}^{\,i}\,\overline{e}{}^{\,
\mu}_{\phantom{a}\hat{i}}(s).
\end{array}
\label{W1}
\end{equation}
Let $\breve{q}{}^{\,r}(\breve{q}{}^{\,0},\, \breve{q}{}^{\,1})$ be
"coordinates relative to the accelerated observer" in the
neighborhood of the accelerated path in MS, with spacetime
components implying
\begin{equation}
\begin{array}{l}
d\breve{q}{}^{\,0}=d\breve{x}{}^{\,0},\quad
d\breve{q}{}^{\,1}=|d\vec{\breve{x}}|,\quad
\vec{\breve{e}}=\frac{d\vec{\breve{x}}}{d\breve{q}{}^{\,1}}=\frac{d\vec{\breve{x}}}{|d\vec{\breve{x}}|},\quad
\vec{\breve{e}}\cdot\vec{\breve{e}}=1.
\end{array}
\label{W02}
\end{equation}
As long as a locality assumption holds, we may describe, with equal
justice, the event at $x^{\mu}$ (\ref{W1}) to be at point
$\breve{q}{}^{\,r}$, such that $x^{\mu}$ and $\breve{q}{}^{\,r}$, in
full generality, are connected via $\breve{q}{}^{\,0}=s$ and
\begin{equation}
\begin{array}{l}
x^{\mu}=z^{\mu}_{q}(s)+\breve{q}{}^{\,1}\,\overline{\beta}{}^{\mu}_{\phantom{a}\hat{1}}(s),
\end{array}
\label{W2}
\end{equation}
where the displacement vector from the origin reads
$dz^{\mu}_{q}(s)=\overline{\beta}{}^{\,\mu}_{\phantom{a}\hat{0}}\,d\breve{q}{}^{\,0}$,
and the components $\overline{\beta}{}^{\,\mu}_{\phantom{a}\hat{r}}$
can be written in terms of
$\overline{e}{}_{\phantom{a}\hat{a}}^{\,\mu}$. Actually, from
(\ref{W1}) and (\ref{W2}) we may obtain
\begin{equation}
\begin{array}{l}

dx^{\mu}=dz^{\mu}_{q}(s)+d\breve{q}{}^{\,1}\,\overline{\beta}{}^{\,\mu}_{\phantom{a}\hat{1}}(s)+
\breve{q}{}^{\,1}\,d\overline{\beta}^{\,\mu}_{\phantom{a}\hat{1}}(s)=
\left[\overline{\beta}{}^{\,\mu}_{\phantom{a}\hat{0}}(1+
\breve{q}{}^{\,1}\check{\varphi}_{0})+\overline{\beta}{}^{\,\mu}_{\phantom{a}\hat{1}}\,\breve{q}{}^{\,1}\check{\varphi}_{1}\right]d\breve{q}{}^{\,0}+
\overline{\beta}{}^{\,\mu}_{\phantom{a}\hat{1}}\,d\breve{q}{}^{\,1}\equiv
\\
dz^{\mu}(s)+d\breve{x}{}^{\,i}\,\overline{e}{}^{\,\mu}_{\phantom{a}\hat{i}}(s)+
\breve{x}{}^{\,i}\,d\overline{e}{}^{\,\mu}_{\phantom{a}\hat{i}}(s)=\left[\overline{e}{}^{\,\mu}_{\phantom{a}\hat{0}}(1+
\breve{x}{}^{\,i}\Phi^{0}_{i})+\overline{e}{}^{\,\mu}_{\phantom{a}\hat{j}}\,\breve{x}{}^{\,i}\Phi^{j}_{i})\right]\,
d\breve{x}{}^{\,0}+
\overline{e}{}^{\,\mu}_{\phantom{a}\hat{i}}\,d\breve{x}{}^{\,i},
\end{array}
\label{W3}
\end{equation}
where $d\overline{\beta}{}^{\,\mu}_{\phantom{a}\hat{1}}(s)$ is
written in the basis
$\overline{\beta}{}^{\,\mu}_{\phantom{a}\hat{a}}$ as
$d\overline{\beta}{}^{\,\mu}_{\phantom{a}\hat{1}}=(\check{\varphi}_{0}\overline{\beta}{}^{\,\mu}_{\phantom{a}\hat{0}}+
\check{\varphi}_{1}\overline{\beta}{}^{\,\mu}_{\phantom{a}\hat{1}})d\breve{q}{}^{\,0}$.
The equation (\ref{W3}) holds if one identifies
\begin{equation}
\begin{array}{l}
 \overline{\beta}{}^{\,\mu}_{\phantom{a}\hat{0}}\left(1+
\breve{q}{}^{\,1}\breve{\varphi}_{0}\right)\equiv
\overline{e}{}^{\,\mu}_{\phantom{a}\hat{0}}\left(1+
\breve{x}{}^{\,i}\Phi^{0}_{i}\right),\quad
\overline{\beta}{}^{\,\mu}_{\phantom{a}\hat{1}}\,\breve{q}{}^{\,1}\breve{\varphi}_{1}\equiv
\overline{e}{}^{\,\mu}_{\phantom{a}\hat{j}}\,\breve{x}{}^{\,i}\Phi^{j}_{i},\quad
\overline{\beta}{}^{\,\mu}_{\phantom{a}\hat{1}}\,d\breve{q}{}^{\,1}\equiv
\overline{e}{}^{\,\mu}_{\phantom{a}\hat{i}}\,d\breve{x}{}^{\,i}.
\end{array}
\label{W4}
\end{equation}
Choosing $\overline{\beta}{}^{\,\mu}_{\phantom{a}\hat{0}}\equiv
\overline{e}{}^{\,\mu}_{\phantom{a}\hat{0}}$, we have then
\begin{equation}
\begin{array}{l}
\breve{q}{}^{\,1}\breve{\varphi}_{0}=
\breve{x}{}^{\,i}\,\Phi^{0}_{i},\quad
\overline{\beta}{}^{\,\mu}_{\phantom{a}\hat{1}}=\overline{e}{}^{\,\mu}_{\phantom{a}\hat{i}}\,\breve{e}{}^{\,i},\quad
\breve{q}{}^{\,1}\breve{\varphi}_{1}
=\breve{x}{}^{\,i}\,\Phi^{j}_{i}\,\breve{e}{}^{\,-1}_{i},
\end{array}
\label{W5}
\end{equation}
with $\breve{e}{}^{j}\,\breve{e}{}^{\,-1}_{i}=\delta^{j}_{i}$.
Consequently, (\ref{W3}) yields the standard metric of
semi-Riemannian 4D background space $V^{(0)}_{4}$, in noninertial
system of the accelerating and rotating observer, computed on the
base of hypothesis of locality:
\begin{equation}
\begin{array}{l}
 \breve{g}=\eta_{\mu\nu}\,dx^{\mu}\otimes dx^{\nu}=\left[(1+
\vec{a}\cdot
\vec{\breve{x}})^{2}+(\vec{\omega}\cdot\vec{\breve{x}})^{2}-(\vec{\omega}\cdot\vec{\omega})(\vec{\breve{x}}\cdot
\vec{\breve{x}})\right]\,d\breve{x}{}^{0}\otimes
d\breve{x}{}^{0}-2\,(\vec{\omega}\wedge\vec{\breve{x}}) \cdot
d\vec{\breve{x}}\otimes d\breve{x}{}^{0}- d\vec{\breve{x}}\otimes
d\vec{\breve{x}},
\end{array}
\label{W19}
\end{equation}
This metric was derived by \cite{Hehl7} and \cite{HL}, in agreement
with \cite{Ni77} -\cite{Ni78D} (see also \cite{MB1, MB2}). We see
that the hypothesis of locality leads to the 2D semi-Riemannian MS
space :\, $V^{(0)}_{2}$ with the incomplete metric
$\breve{g}\quad(\varrho=0)$:
\begin{equation}
\begin{array}{l}
\breve{g}=\left[(1+
\breve{q}{}^{\,1}\breve{\varphi}_{0})^{2}-(\breve{q}{}^{\,1}\breve{\varphi}_{1})^{2}\right]\,d\breve{q}{}^{\,0}\otimes
d\breve{q}{}^{\,0}-  2\,(\breve{q}{}^{\,1}\breve{\varphi}_{1})
d\breve{q}{}^{\,1}\otimes d\breve{q}{}^{\,0}-
d\breve{q}{}^{\,1}\otimes d\breve{q}{}^{\,1},
\end{array}
\label{W05}
\end{equation}
Therefore, our strategy now is to deform the metric (\ref{W05}) by
an additional deformation of semi-Riemannian 4D background space
$V^{(0)}_{4}\rightarrow \widetilde{\mathcal{M}}_{4}\equiv
V^{(\varrho)}_{4}$,  which, as a corollary, will recover the
complete metric $g\quad(\varrho\neq 0)$ (\ref{Tau2}) of the
distorted MS\,-\, $V^{(\varrho)}_{2}$. Following \cite{gago1}, this
means that we should find the first deformation matrices,
$\pi(\varrho):\phantom{a}=(\pi^{\phantom{a}\hat{b}}_{\hat{a}})(\varrho)$,
which yield the local tetrad deformations
\begin{equation}
\begin{array}{l}
e_{\hat{c}}=\pi^{\phantom{a}\hat{a}}_{\hat{c}}\,\breve{e}_{\hat{a}},\quad
\vartheta^{\hat{c}}=\pi_{\phantom{a}\hat{b}}^{\hat{c}}\,\breve{\vartheta}{}^{\hat{b}},\quad
e\,\vartheta=e_{\hat{a}}\otimes\vartheta^{\hat{a}}=\Omega^{a}_{\phantom{a}\hat{b}}\,
\breve{e}_{\hat{a}}\otimes\breve{\vartheta}{}^{\hat{b}},
\end{array}
\label{R30}
\end{equation}
where $\Omega_{\phantom{a}\hat{b}}^{\hat{a}}(\varrho)=
\pi^{\phantom{a}\hat{a}}_{\hat{c}}(\varrho)\,\pi^{\phantom{a}\hat{c}}_{\hat{b}}(\varrho)$
is referred to as the anholonomic {\em deformation tensor},
 and that the resulting deformed metric of the space
$V^{(\varrho)}_{4}$ can be split as
\begin{equation}
\begin{array}{l}
g_{\mu\nu}(\varrho)=\Upsilon^{2}(\varrho)\,\breve{g}_{\mu\nu}+\gamma_{\mu\nu}(\varrho),
\end{array}
\label{R38}
\end{equation}
provided
\begin{equation}
\begin{array}{l}
\gamma_{\mu\nu}(\varrho)=[\gamma_{\hat{a}\hat{b}}-\Upsilon^{2}(\varrho)\,o_{\hat{a}\hat{b}}]\,
\breve{e}{}^{\,\hat{a}}_{\phantom{a}\mu}\,\breve{e}{}^{\,\hat{b}}_{\phantom{a}\nu},\quad
\gamma_{\hat{c}\hat{d}}=o_{\hat{a}\hat{b}}\,\pi^{\phantom{a}\hat{a}}_{\hat{c}}\,\pi_{\hat{d}}^{\phantom{a}\hat{b}},
\end{array}
\label{W20}
\end{equation}
where $\Upsilon(\varrho)=\pi^{\hat{a}}_{\hat{a}}(\varrho)$ and
$\gamma_{\hat{a}\hat{b}}(\breve{x})$ are the second deformation
matrices. Let the Latin letters  $\hat{r},\hat{s},...=0,1$ be the
anholonomic indices related to the anholonomic frame
$e_{\hat{r}}=e_{\phantom{a}\hat{r}}^{s}\,\partial_{\tilde{s}}$,
defined on the $V^{(\varrho)}_{2}$, with
$\partial_{\tilde{s}}=\partial/\partial\,\widetilde{q}{}^{\tilde{s}}$
as the vectors tangent to the coordinate lines. So, a smooth
differential 2D-manifold $V^{(\varrho)}_{2}$ has at each point
$\widetilde{q}^{s}$ a tangent space
$\widetilde{T}_{\widetilde{q}}V^{(\varrho)}_{2}$, spanned by the
frame, $\{e_{\hat{r}}\}$, and the coframe members
$\vartheta^{\hat{r}}=e^{\phantom{a}\hat{r}}_{s}\,d\widetilde{q}{}^{\tilde{s}}$,
which constitute a basis of the covector space
$\widetilde{T}{}^\star_{\widetilde{q}}V^{(\varrho)}_{2}$. All this
nomenclature can be given for $V^{(0)}_{2}$ too. Then, we may
calculate corresponding vierbein fields
$\breve{e}{}^{\phantom{a}\hat{s}}_{r}$ and
$e^{\phantom{a}\hat{s}}_{r}$  from the equations
\begin{equation}
\begin{array}{l}
\breve{g}_{rs}=\breve{e}{}^{\phantom{a}\hat{r'}}_{r}\,\breve{e}{}^{\phantom{a}\hat{s'}}_{s}\,o_{\hat{r'}\hat{s'}},\quad
g_{\tilde{r}\tilde{s}}=e^{\phantom{a}\hat{r'}}_{r}\,e^{\phantom{a}\hat{s'}}_{s}\,o_{\hat{r'}\hat{s'}},
\end{array}
\label{W09}
\end{equation}
with  $\breve{g}_{rs}$ and $g_{\tilde{r}\tilde{s}}$  given by
(\ref{W05}) and (\ref{RE494}), respectively. Hence
\begin{equation}
\begin{array}{l}
\breve{e}{}^{\phantom{a}\hat{0}}_{0}=1+ \vec{a}\cdot
\vec{\breve{x}},\quad
\breve{e}{}^{\phantom{a}\hat{1}}_{0}=\vec{\omega}\cdot\vec{\breve{x}},\quad
\breve{e}{}^{\phantom{a}\hat{0}}_{1}=0,\quad
\breve{e}{}^{\phantom{a}\hat{1}}_{1}=1,\\
e^{\phantom{a}\hat{0}}_{0}=1+\frac{\varrho v_{q}}{\sqrt{2}},\quad
e^{\phantom{a}\hat{1}}_{0}=\frac{\varrho}{\sqrt{2}},\quad
e^{\phantom{a}\hat{0}}_{1}=-\frac{\varrho}{\sqrt{2}},\quad
e^{\phantom{a}\hat{1}}_{1}=1-\frac{\varrho v_{q}}{\sqrt{2}}.
\end{array}
\label{W10}
\end{equation}
Since a distortion of MS may affect only the MS-part of the
components $\overline{\beta}{}^{\,\mu}_{\phantom{a}\hat{r}}$,
without relation to the background spacetime part, therefore, a
deformation $V^{(0)}_{4}\rightarrow V^{(\varrho)}_{4}$  is
equivalent to a straightforward generalization
$\overline{\beta}{}^{\,\mu}_{\phantom{a}\hat{r}}\rightarrow
\beta^{\mu}_{\phantom{a}\hat{r}}$, where
\begin{equation}
\begin{array}{l}
\beta^{\mu}_{\phantom{a}\hat{r}}=
E^{\phantom{a}\hat{s}}_{\hat{r}}\,\overline{\beta}{}^{\,\mu}_{\phantom{a}\hat{s}},\quad
E^{\phantom{a}\hat{s}}_{\hat{r}}:\phantom{a}=
e^{r'}_{\phantom{a}\hat{r}}\,\breve{e}{}^{\phantom{a}\hat{s}}_{r'}.
\end{array}
\label{W6}
\end{equation}
Consequently, the (\ref{W6}) gives a generalization of (\ref{W1}) as
\begin{equation}
\begin{array}{l}
x^{\mu}\,\rightarrow\,x^{\mu}_{(\varrho)}=z^{\mu}_{(\varrho)}(s)+
\breve{x}{}^{\,i}\,e^{\mu}_{\phantom{a}\hat{i}}(s),
\end{array}
\label{W8}
\end{equation}
provided, as before, $\breve{x}{}^{\mu}$ denotes the coordinates
relative to the accelerated observer in 4D background space
$V_{4}^{(\varrho)}$, and according to (\ref{W4}), we have
\begin{equation}
\begin{array}{l}
e^{\mu}_{\phantom{a}\hat{0}}=\beta^{\mu}_{\phantom{a}\hat{0}},\quad
e^{\mu}_{\phantom{a}\hat{i}}=\beta^{\mu}_{\phantom{a}\hat{i}}\,\breve{e}{}^{-1}_{i}.
\end{array}
\label{W9}
\end{equation}
A displacement vector from the origin is then
$dz^{\mu}_{\varrho}(s)=e^{\mu}_{\phantom{a}\hat{0}}\,d\breve{x}{}^{0}$,
Combining (\ref{W6}) and (\ref{W9}), and inverting
$e^{\phantom{a}\hat{s}}_{r}$ (\ref{W10}), we obtain  $
e^{\mu}_{\phantom{a}\hat{a}}=\pi^{\phantom{a}\hat{b}}_{\hat{a}}(\varrho)\,\overline{e}{}^{\,\mu}_{\phantom{a}\hat{b}}$,
where
\begin{equation}
\begin{array}{l}
 \pi^{\hat{0}}_{\hat{0}}(\varrho)\equiv
(1+\frac{\varrho^{2}}{2\gamma^{2}_{q}})^{-1}(1-\frac{\varrho
v_{q}}{\sqrt{2}})\,(1+ \vec{a}\cdot \vec{\breve{x}}),\quad
\pi^{\hat{i}}_{\hat{0}}(\varrho)\equiv
-(1+\frac{\varrho^{2}}{2\gamma^{2}_{q}})^{-1}\frac{\varrho}{\sqrt{2}}\,\breve{e}^{i}\,(1+
\vec{a}\cdot
\vec{\breve{x}}),\\

\pi^{\hat{0}}_{\hat{i}}(\varrho)\equiv(1+\frac{\varrho^{2}}{2\gamma^{2}_{q}})^{-1}\left[(\vec{\omega}\cdot\vec{\breve{x}})(1-\frac{\varrho
v_{q}}{\sqrt{2}})-\frac{\varrho}{\sqrt{2}}\right]\,\breve{e}^{-1}_{i},\quad
\pi^{\hat{j}}_{\hat{i}}(\varrho)=\delta^{j}_{i}\,\pi(\varrho),\\
\pi(\varrho)\equiv(1+\frac{\varrho^{2}}{2\gamma^{2}_{q}})^{-1}\,\left[(\vec{\omega}\cdot\vec{\breve{x}})\,\frac{\varrho}{\sqrt{2}}+
1+\frac{\varrho v_{q}}{\sqrt{2}}\right] .
\end{array}
\label{W12}
\end{equation}
Thus,
\begin{equation}
\begin{array}{l}
dx^{\mu}_{\varrho}=dz^{\mu}_{\varrho}(s)+d\breve{x}{}^{\,i}\,e^{\,\mu}_{\phantom{a}\hat{i}}+
\breve{x}{}^{\,i}\,de^{\,\mu}_{\phantom{a}\hat{i}}(s)=(\tau^{\hat{b}}\,d\breve{x}{}^{0}+
\pi^{\hat{b}}_{\hat{i}}\,d\breve{x}{}^{\,i})\,\overline{e}{}^{\,\mu}_{\phantom{a}\hat{b}}\,,
\end{array}
\label{W13}
\end{equation}
where
\begin{equation}
\begin{array}{l}
\tau^{\hat{b}}\equiv
\pi^{\hat{b}}_{\hat{0}}+\breve{x}{}^{\,i}\,\left(\pi^{\hat{a}}_{\hat{i}}\Phi^{b}_{a}+\frac{d\pi^{\hat{b}}_{\hat{i}}}{ds}\right).
\end{array}
\label{W14}
\end{equation}
Hence, in general, the metric in noninertial frame of arbitrary
accelerating and rotating observer in Minkowski spacetime  is
\begin{equation}
\begin{array}{l}
g(\varrho)=\eta_{\mu\nu}\,dx^{\mu}_{\varrho}\otimes
dx^{\nu}_{\varrho}= W_{\mu\nu}(\varrho)\,d\breve{x}{}^{\mu}\otimes
d\breve{x}^{\nu},
\end{array}
\label{W15}
\end{equation}
which can be conveniently decomposed according to
\begin{equation}
\begin{array}{l}
W_{00}(\varrho)= \pi^{2}\left[(1+ \vec{a}\cdot
\vec{\breve{x}})^{2}+(\vec{\omega}\cdot\vec{\breve{x}})^{2}-(\vec{\omega}\cdot\vec{\omega})(\vec{\breve{x}}\cdot
\vec{\breve{x}})\right]+\gamma_{00}(\varrho),\\
 W_{0i}(\varrho)=-\pi^{2}\,(\vec{\omega}\wedge\vec{\breve{x}})^{i}+\gamma_{0i}(\varrho),\quad
W_{ij}(\varrho)=-\pi^{2}\,\delta_{ij}+\gamma_{ij}(\varrho),
\end{array}
\label{W16}
\end{equation}
and that
\begin{equation}
\begin{array}{l}
 \gamma_{00}(\varrho)=\pi\,\left[(1+ \vec{a}\cdot
\vec{\breve{x}})\zeta^{0}-
(\vec{\omega}\wedge\vec{\breve{x}})\cdot\vec{\zeta}\right]+(\zeta^{0})^{2}-(\vec{\zeta})^{2},\quad
 \gamma_{0i}(\varrho)=-\pi\,\zeta^{i}+\tau^{\hat{0}}\,\pi^{\hat{0}}_{\hat{i}},\\

\gamma_{ij}(\varrho)=\pi^{\hat{0}}_{\hat{i}}\,\pi^{\hat{0}}_{\hat{j}},\quad
\zeta^{0}= \pi\,\left(\tau^{\hat{0}}-1- \vec{a}\cdot
\vec{\breve{x}}\right),\quad
\vec{\zeta}=\pi\,\left(\vec{\tau}-\vec{\omega}\wedge\vec{\breve{x}}\right).
\end{array}
\label{W17}
\end{equation}
As we expected, according to (\ref{W15})- (\ref{W17}), the matric
$g(\varrho)$ is decomposed in the form of (\ref{R38}):
\begin{equation}
\begin{array}{l}
g(\varrho)=\pi^{2}(\varrho)\,\breve{g} +\gamma(\varrho),
\end{array}
\label{W015}
\end{equation}
where
$\gamma(\varrho)=\gamma_{\mu\nu}(\varrho)\,d\breve{x}{}^{\mu}\otimes
d\breve{x}^{\nu}$ and
$\Upsilon(\varrho)=\pi^{\hat{a}}_{\hat{a}}(\varrho)= \pi(\varrho)$,
provided (\ref{RL888}) gives $(s=s_{q})$
\begin{equation}
\begin{array}{l}
\frac{\tau_{2}\varrho}{\sqrt{2}}=\int^{s}_{0}|{\bf a}|ds'.
\end{array}
\label{W7}
\end{equation}
In general, the geodesic coordinates are admissible as long as
\begin{equation}
\begin{array}{l}
\left(1+ \vec{a}\cdot
\vec{\breve{x}}+\frac{\zeta^{0}}{\pi}\right)^{2}\,>\,\left(\vec{\omega}\wedge\vec{\breve{x}}+\frac{\vec{\zeta}}{\pi}\right)^{2}.
\end{array}
\label{W22}
\end{equation}
The equations (\ref{W19}) and (\ref{W15}) say that the vierbein
fields, with entries
$\eta_{\mu\nu}\,\overline{e}{}^{\,\mu}_{\phantom{a}\hat{a}}\,\overline{e}^{\,\nu}_{\phantom{a}\hat{b}}=o_{\hat{a}\hat{b}}$,
lead to the relations
\begin{equation}
\begin{array}{l}
\breve{g}=o_{\hat{a}\hat{b}}\,\breve{\vartheta}{}^{\hat{a}}\otimes\breve{\vartheta}{}^{\hat{b}},\quad
g=o_{\hat{a}\hat{b}}\,\vartheta^{\hat{a}}\otimes\vartheta^{\hat{b}},
\end{array}
\label{W012}
\end{equation}
and that (\ref{W3}) and (\ref{W13}) readily give the coframe fields:
\begin{equation}
\begin{array}{l}
\breve{\vartheta}{}^{\hat{b}}=\overline{e}{}_{\,\mu}^{\phantom{a}\hat{b}}\,dx^{\mu}=
\breve{e}{}^{\hat{b}}_{\phantom{a}\mu}\,d\breve{x}{}^{\,\mu},\quad
\breve{e}{}^{\hat{b}}_{\phantom{a}0}=N^{b}_{0},\quad
\breve{e}{}^{\hat{b}}_{\phantom{a}i}=N^{b}_{i},\\
\vartheta^{\hat{b}}=\overline{e}{}_{\,\mu}^{\phantom{a}\hat{b}}\,dx^{\mu}_{\varrho}=
e^{\hat{b}}_{\phantom{a}\mu}\,d\breve{x}{}^{\,\mu}=\pi^{\hat{b}}_{\phantom{a}\hat{a}}\,\breve{\vartheta}{}^{\hat{a}},\quad
e^{\hat{b}}_{\phantom{a}0}=\tau^{\hat{b}},\quad
e^{\hat{b}}_{\phantom{a}i}=\pi^{\hat{b}}_{\phantom{a}\hat{i}}.
\end{array}
\label{W24}
\end{equation}
where $N^{0}_{0}=N\equiv \left(1+ \vec{a}\cdot
\vec{\breve{x}}\right),\quad N^{0}_{i}=0,\quad N^{i}_{0}=N^{i}\equiv
\left(\vec{\omega}\cdot\vec{\breve{x}}\right)^{i},\quad
N^{j}_{i}=\delta^{j}_{i}$. In the standard $(3+1)$-decomposition of
spacetime, $N$ and $N^{i}$ are known as {\it lapse function} and
{\it shift vector}, respectively \cite{Gron}. Hence, we may easily
recover the frame field
$e_{\hat{a}}=e_{\hat{b}}^{\phantom{a}\mu}\,\breve{e}_{\mu}=\pi^{\phantom{a}\hat{b}}_{\hat{a}}\,\breve{e}_{\hat{b}}$\,\,
by inverting (\ref{W24}):
\begin{equation}
\begin{array}{l}
e_{\hat{0}}=\frac{\pi}{\pi\,\tau^{\hat{0}}-\pi^{\hat{0}}_{\hat{k}}\,\tau^{\hat{k}}}\,\breve{e}_{0}\,-\,
\frac{\tau^{\hat{i}}}
{\pi\,\tau^{\hat{0}}-\pi^{\hat{0}}_{\hat{k}}\,\tau^{\hat{k}}}\,\breve{e}_{i},\quad
e_{\hat{i}}=-\frac{\pi^{\hat{0}}_{\hat{i}}}
{\pi\,\tau^{\hat{0}}-\pi^{\hat{0}}_{\hat{k}}\,\tau^{\hat{k}}}\,\breve{e}_{0}\,+\pi^{-1}\left[\delta^{j}_{i}+
\frac{\tau^{j}\,\pi^{\hat{0}}_{\hat{i}}}
{\pi\,\tau^{\hat{0}}-\pi^{\hat{0}}_{\hat{k}}\,\tau^{\hat{k}}}\right]\,\breve{e}_{j}.
\end{array}
\label{W25}
\end{equation}

\section{Involving the background semi-Riemann space $V_{4}$; Justification for
the introduction of the PE} We can always choose {\em natural
coordinates} $X^{\alpha}(T,X,Y,Z)=(T,\,\vec{X})$ with respect to the
axes of the local free-fall coordinate frame $S_{4}^{(l)}$ in an
immediate neighbourhood of any spacetime point $(\breve{x}_{p})\in
V_{4}$ in question of the background semi- Riemann space, $V_{4}$,
over a differential region taken small enough so that we can neglect
the spatial and temporal variations of gravity for the range
involved. The values of the metric tensor $\breve{g}_{\mu\nu}$ and
the affine connection $\breve{\Gamma}^{\lambda}_{\mu\nu}$ at the
point $(\breve{x}_{p})$ are necessarily sufficient information for
determination of the natural coordinates
$X^{\alpha}(\breve{x}{}^{\mu})$ in the small region of the
neighbourhood of the selected point~\cite{W72}. Then the whole
scheme outlined in the section 4 will be held in the frame
$S_{4}^{(l)}$. The relativistic gravitational force
$\breve{f}{}^{\mu}_{g}(\breve{x})$ exerted on the test particle of
the mass $(m)$ is given by
\begin{equation}
\begin{array}{l}
\breve{f}{}^{\mu}_{g}(\breve{x})=m\frac{d^{2}\breve{x}{}^{\mu}}{d\breve{s}^{2}}=-
m\breve{\Gamma}{}^{\mu}_{\nu\lambda}(a)\frac{d\breve{x}{}^{\nu}}{d\breve{s}}
\frac{d\breve{x}^{\lambda}}{d\breve{s}},
\end{array}
\label{RG1}
\end{equation}
such that the gravitational force in the free-fall coordinate frame
$S_{4}^{(l)}$ will be
\begin{equation}
\begin{array}{l}
f^{\alpha}_{g(l)}=\frac{\partial X^{\alpha}}{\partial
\breve{x}{}^{\mu}}\breve{f}{}^{\mu}_{g},
\end{array}
\label{RG15}
\end{equation}
As before, the two systems $S_{2}$ and $S_{4}^{(l)}$ can be chosen
in such a way as the axis $\vec{e}_{q}$ of $S_{(2)}$ lies
($\vec{e}_{q} =\vec{e}_{f}$) along the acting net force $\vec{f}=
\vec{f}_{(l)}+\vec{f}_{g(l)}$, where $\vec{f}_{(l)}$ is the SR value
of the unbalanced relativistic force other than gravitational in the
frame $S_{4}^{(l)}$, while the time coordinates in the two systems
are taken the same, $q^{0}=t=X^{0}=T.$ The (\ref{RLL9}) now can be
replaced by
\begin{equation}
\begin{array}{l}
\frac{1}{\sqrt{2}}\frac{d(\tau_{2}\varrho)}{ds_{q}}=
\frac{1}{m}|f^{\alpha}_{(l)}+f^{\alpha}_{g(l)}|,
\end{array}
\label{RLL12G}
\end{equation}
and according to (\ref{G33}), the  general {\em inertial force}
reads
\begin{equation}
\begin{array}{l}
\breve{\vec{f}}_{(in)}=m\vec{a}_{in}=
-\frac{m\vec{a}_{abs}}{\gamma_{q}}=
-\frac{\vec{e}_{f}}{\gamma_{q}}|f^{\alpha}_{(l)}- m\frac{\partial
X^{\alpha}}{\partial
\breve{x}{}^{\sigma}}\breve{\Gamma}{}^{\sigma}_{\mu\nu}
\frac{d\breve{x}{}^{\mu}}{dS} \frac{d\breve{x}^{\nu}}{dS}|.
\end{array}
\label{RLLL9}
\end{equation}
Despite of totally different and independent sources of gravitation
and inertia, at $f^{\alpha}_{(l)}=0$, the (\ref{RLLL9}) establishes
the independence of free-fall trajectories of the mass,  internal
composition and structure of bodies. This furnishes a justification
for the introduction of the PE. A remarkable feature is that,
although the inertial force has a nature different than the
gravitational force, nevertheless both are due to a distortion of
the local inertial properties of, respectively,  2D MS and
4D-background space. The non-vanishing inertial force acting on the
photon of energy $h\nu$, and that of effective mass
$\left(h\nu/c^{2}\right)$, after inserting units $(h,\,c)$ which so
far was suppressed, can be obtained from the ~(\ref{RLLL9})
($f^{\alpha}_{(l)}=0$) as
\begin{equation}
\begin{array}{l}
\breve{\vec{f}}_{(in)}= -\left(\frac{h\nu}{c^{2}}\right)\vec{e}_{f}
|\frac{\partial X^{\alpha}}{\partial
\breve{x}{}^{\sigma}}\Gamma^{\sigma}_{\mu\nu}
\frac{d\breve{x}{}^{\mu}}{dT} \frac{d\breve{x}^{\nu}}{dT}|=
-\left(\frac{h\nu}{
c^{2}}\right)\vec{e}_{f}|(\frac{d^{2}\widetilde{t}}{dT^{2}})
\frac{dX^{\alpha}}{d\widetilde{t}}+(\frac{d\widetilde{t}}{dT})^{2}\frac{\partial
X^{\alpha}}{\partial \breve{x}{}^{\,i}}\frac{d
u_{i}}{d\widetilde{t}}|,
\end{array}
\label{RLLL10}
\end{equation}
provided $\vec{e}_{f}=(\vec{X}/|\vec{X}|)$, $v_{q}=(\vec{e}_{f}\cdot
\breve{\vec{u}})=|\breve{\vec{u}}|,\,\,(\gamma_{q}=\gamma)$ where
$\breve{\vec{u}}$ is the velocity of a photon and
 $(d \breve{\vec{u}}/d\widetilde{t})$ is the
acceleration, and that,
$\breve{g}_{\mu\nu}(d\breve{x}{}^{\mu}/dT)\otimes
(d\breve{x}^{\nu}/dT)=0.$ To obtain some feeling for this, in the
(PPN) approximation~\cite{will1}-\cite{will3} we may calculate the
inertial force exerted on the photon \cite{gago}, in a gravitating
system of particles that are bound together by their mutual
gravitational attraction to order $\bar{v}^{2}\sim
G_{N}\bar{M}/\bar{r}$ of a small parameter, where
$\bar{v},\,\bar{M}$ and $\bar{r}$ are typically the average values
of their velocities, masses and separations, respectively. To this
aim, we may expand the metric tensor to the following order: $
\breve{g}_{00}=1+\stackrel{2}{g}_{00}+\stackrel{4}{g}_{00}+...,\quad
\breve{g}_{ij}=-\delta_{ij}+\stackrel{2}{g}_{ij}+\stackrel{4}{g}_{ij}+...,\quad
 \breve{g}_{i0}=\stackrel{3}{g}_{i0}+\stackrel{5}{g}_{i0}+....,
$ where $\stackrel{N}{g}_{\mu\nu}$ denotes the term of order
$\bar{v}^{N}.$ Taking into account the standard  expansions of the
affine connection ~\cite{W72}:\,
$\breve{\Gamma}^{\sigma}_{\mu\nu}=\stackrel{2}{\Gamma^{\sigma}_{\mu\nu}}+\stackrel{4}{\Gamma^{\sigma}_{\mu\nu}}+...$
for the components
$\breve{\Gamma}{}^{i}_{00},\,\breve{\Gamma}{}^{i}_{jk},\,
\breve{\Gamma}{}^{0}_{0i},$ and that
$\breve{\Gamma}^{\sigma}_{\mu\nu}=\stackrel{3}{\Gamma^{\sigma}_{\mu\nu}}+\stackrel{5}{\Gamma^{\sigma}_{\mu\nu}}+...$
for the components
$\breve{\Gamma}^{i}_{0j},\,\breve{\Gamma}{}^{0}_{00},\,
\breve{\Gamma}{}^{0}_{ij},$ where
$\stackrel{2}{\Gamma^{i}_{00}}=\stackrel{2}{\Gamma^{0}_{0i}}
=-(1/2)(\partial\stackrel{2}{g}_{00}/\partial \breve{x}{}^{\,i})$
etc., hence to the required accuracy we obtain
\begin{equation}
\begin{array}{l}
\breve{\vec{f}}{}^{(2)}_{(in)}=
-\left(\frac{h\nu}{c^{2}}\right)\vec{e}_{f}
|\stackrel{1}{(\frac{\partial X^{\alpha}}{\partial
\breve{x}{}^{\sigma}})}
\stackrel{2}{(\frac{d^{2}\breve{x}{}^{\sigma}}{dT^{2}})}|=
-\left(\frac{h\nu}{c^{2}}\right)
\stackrel{2}{(\frac{d\breve{\vec{u}}}{d\widetilde{t}})}=
-\left(\frac{h\nu}{\gamma
c^{2}}\right)[-2\vec{\nabla}\phi+4\breve{\vec{u}}(\breve{\vec{u}}\cdot
\vec{\nabla}\phi) +   O(\bar{v}{}^{3})],
\end{array}
\label{RG2}
\end{equation}
where $\phi$ is the Newton potential, such that
$\stackrel{2}{g}_{00}=2\phi,$ $
\stackrel{2}{g}_{ij}=2\delta_{ij}\phi,$ and
$|\breve{\vec{u}}|=1+2\phi+O(\bar{v}{}^{3}).$

\section{RTI in the background post Riemannian geometry}
Recall that the general metric-affine space,
$(\widetilde{\mathcal{M}}_{4},$ $\,g,\,\Gamma)$, is defined to have
equipped with two independent geometrical structures: the
pseudo-Riemannian metric, $g$ and the linear affine connection
$\Gamma$. The new geometrical property of the spacetime, are the
{\em nonmetricity} 1-form $N_{ab}$ and the affine {\em torsion}
2-form $T^{a}$ representing a translational misfit (for a
comprehensive discussion see~\cite{Hehl3}-\cite{Popl}. These,
together with the {\em curvature} 2-form $R_{a}^{\phantom{a}b}$,
symbolically can be presented as $
\left(N_{ab},\,T^{a},\,R_{a}^{\phantom{a}b}\right)\;\sim\; {\cal
D}\left(g_{ab},\,\vartheta^{a},\, \Gamma_{a}^{\phantom{a}b}\right),
$ where for a tensor-valued $p-$form density of representation type
$\rho(L^{b}_{\phantom{a} a})$, the $GL(4,R)$-covariant exterior
derivative reads ${\cal D}:\phantom{a} =d+\Gamma^{\phantom{a}
b}_{a}\rho(L^{b}_{\phantom{a} a})\,\wedge$\,. To avoid any
possibility of confusion, here and throughout we use the first half
of Latin alphabet ($a, b, c, . . . = 0,1, 2, 3$ rather than $(\pm)$)
now to denote the anholonomic indices related to the tangent space,
which is endowed with the Lorentzian metric $o_{ab}:\,=diag(+ - -
-)$. If the nonmetricity tensor $N_{\lambda\mu\nu}=-{\cal
D}_{\lambda}\,g_{\mu\nu}\equiv -g_{\mu\nu\,; \lambda}$ does not
vanish, the general formula for the affine connection written in the
spacetime components is~\cite{Popl}
\begin{equation}
\begin{array}{l}
\Gamma^{\rho}_{\phantom{a}\mu\,\nu}=\stackrel{\circ
}{\Gamma}{}^{\rho}_{\phantom{a}\mu\,\nu}
+K^{\rho}_{\phantom{k}\mu\nu}-N^{\rho}_{\phantom{k}\mu\nu}+\frac{1}{2}N_{(\mu\phantom{k}\nu)}^{\phantom{(i}\rho},
\end{array}
\label{RU1}
\end{equation}
where the metric alone determines the torsion-free  Levi-Civita
connection $\stackrel{\circ}{\Gamma}{}^{\rho}_{\phantom{a}\mu \nu}$,
$K^{\rho}_{\phantom{k}\mu\nu}:\phantom{a}=2Q_{(\mu\nu)}^{\phantom{(ij)}\rho}+
Q^{\rho}_{\phantom{k}\mu\nu}$ is the non-Riemann part - the affine
{\em contortion tensor}. The torsion, $Q^{\rho}_{\phantom{k}\mu\nu}=
\frac{1}{2}\,T^{\rho}_{\phantom{k}\mu\nu}=\Gamma^{\rho}_{\phantom{a}[\mu\,\nu]}$
given with respect to a holonomic frame, $d\,\vartheta^{\rho}=0$, is
a third-rank tensor, antisymmetric in the first two indices, with 24
independent components.

\subsection{The principle of equivalence in the RC space}
The RC manifold, $U_{4}$, is a particular case of general
metric-affine manifold $\widetilde{M}_{4}$, restricted by the
metricity condition $N_{ab}=0$, when a nonsymmetric linear
connection, $\Gamma$, is said to be metric compatible. The space,
$U_{4}$, also locally has the structure of $M_{4}$, as has been
first pointed out by ~\cite{Heyd} and developed
by~\cite{Hehl5}-\cite{Hart}. In the case of the RC space there also
exist orthonormal reference frames which realize an `anholonomic'
free-fall elevator. In Hartley's formulation~\cite{Hart}, this
reads: {\it For
  any single point $P\in U_{4}$, there exist coordinates $\{
  x{}^{\mu}\} $ and an orthonormal frame $\{ e_{a}\} $
  in a neighborhood of $P$ such that
  $$
  \left\{\begin{array}{rcl}
      e_{a}&=&\delta _{a}^{\mu}\,\partial _{x{}^{\mu}}\\ \Gamma
      _{a}^{\phantom{a}b}&=&0
         \end{array}
       \right\}\quad {\rm at}\,\,\, P$$
       where
       $\Gamma_{a}^{\phantom{a}b}$ are the connection 1-forms
referred to the frame $\{e_{a}\}$}. Therefore the existence of
torsion does not violate the PE.  Note that, since $\nabla\,{\bf
g}=0$  holds in $U_{4}$, the arguments showing that ${\bf g}$ can be
transformed to $o$ at any point $P$ in $U_{4}$ are the same as in
the case of $V_{4}$, while the treatment of the connection must be
different: the antisymmetric part of $\omega$ can be eliminated only
by a suitable choice for the relative orientation of neighbouring
tetrads. Actually, let us choose new local coordinates at $P$,
$\phantom{a}d\,x{}^{\mu}\rightarrow d\,x{}^{a} =
e^{a}_{\mu}\,d\,x^{\mu}$, related to an inertial frame. Then,
\begin{equation}
\begin{array}{l}

g_{ab}'=e_{a}^{\phantom{a}\mu}\,e_{b}^{\phantom{a}\nu}\,g_{\mu\nu}=o_{ab},\quad
\Gamma^{\prime
b}_{\phantom{a}ac}=e^{b}_{\phantom{a}\mu}\,e_{a}^{\phantom{a}\nu}\,e_{c}^{\phantom{a}\lambda}\,
(\Delta^{\mu}_{\phantom{a}\nu\lambda}+K^{\mu}_{\phantom{a}\nu\lambda})\equiv
e_{c}^{\phantom{a}\lambda}\,\omega^{b}_{\phantom{a}a\lambda}.
\end{array}
\label{I1}
\end{equation}
As it is argued in~\cite{Blag}, the metricity condition ensures that
this can be done consistently at every point in spacetime. Suppose
that we have a tetrad $\{e_{a}(x)\}$ at the point $P$, and a tetrad
$\{e_{a}(x + d\,x)\}$ at another point in a neighbourhood of $P$;
then, we can apply a suitable Lorentz rotation to $e_{a}(x + d\,x)$,
so that it becomes parallel to $e_{a}(x)$ . Given a vector $v$ at
$P$, it follows that the components $v_{c}=v\cdot e_{c}$ do not
change under parallel transport from $x$ to $x + d\,x$, provided the
metricity condition holds. Hence, the connection coefficients
$\omega^{ab}_{\phantom{ab}\mu}(x)$ at $P$, defined with respect to
this particular tetrad field, vanish:
$\omega^{ab}_{\phantom{ab}\mu}(P)=0$. This property is compatible
with $g'_{ab}=o_{ab}$, since Lorentz rotation does not influence the
value of the metric at a given point. In more general geometries,
where the symmetry of the tangent space is higher than the Poincare
group, the usual form of the PE is violated and local physics
differs from SR.

\subsection{The generalized inertial force exerted on the  extended spinning body in the $U_{4}$}
We now compute the relativistic inertial force for the motion of the
matter, which is distributed over a small region in the $U_{4}$
space and consists of points with the coordinates $x^{\mu}$, forming
an extended body whose motion in the space, $U_{4}$, is represented
by a world tube in spacetime. Suppose the motion of the body as a
whole is represented by an arbitrary timelike world line $\gamma$
inside the world tube, which consists of points with the coordinates
$\tilde{X}{}^{\mu}(\tau)$, where $\tau$ is the proper time on
$\gamma$. Define
\begin{equation}
\begin{array}{l}
\delta x{}^{\mu}=x{}^{\mu}-\tilde{X}{}^{\mu},\phantom{a}\delta
x^{0}=0,\phantom{a}u^{\mu}=\frac{d\,\tilde{X}{}^{\mu}}{d\,s}.
\end{array}
\label{I4}
\end{equation}
The {\em Papapetrou equation of motion for the modified momentum}
(\cite{Pa}-\cite{Mol2}, \cite{Popl}) is
\begin{equation}
\begin{array}{l}
\frac{\stackrel{\circ }{\cal D}\,\Theta^{\nu}}{{\cal
D}\,s}=-\frac{1}{2}\,\stackrel{\circ
}{R}{}^{\nu}_{\phantom{j}\mu\sigma\rho}\,u{}^{\mu}\,
J^{\sigma\rho}-\frac{1}{2}\,N_{\mu\rho\lambda}\,K^{\mu\rho\lambda:\,\nu},
\end{array}
\label{I8}
\end{equation}
where $K^{\mu}_{\nu\lambda}$ is the contortion tensor,
\begin{equation}
\begin{array}{l}
\Theta^{\nu}=P^{\nu}+\frac{1}{u{}^{0}}\,\stackrel{\circ
}{\Gamma}{}^{\phantom{a}\nu}_{\mu\,\rho}\,(u{}^{\mu} \,J^{\rho
0}+N^{0 \mu\rho})-
\frac{1}{2u{}^{0}}\,K_{\mu\rho}^{\phantom{ik}\nu}\, N^{\mu\rho 0}
\end{array}
\label{I7}
\end{equation}
is referred to as the {\em modified 4-momentum},
$P^{\lambda}=\int\tau^{\lambda 0}\,d\,\Omega$ is the ordinary
4-momentum, $d\,\Omega:=d\,x^{4}$, and the following integrals are
defined:
\begin{equation}
\begin{array}{l}
 M^{\mu\rho}=u{}^{0}\,\int\tau^{\mu\rho}\,d\,\Omega, \quad
M^{\mu\nu\rho}=-u{}^{0}\,\int\delta
x{}^{\mu}\,\tau^{\nu\rho}\,d\,\Omega, \quad
N^{\mu\nu\rho}=u{}^{0}\,\int s^{\mu\nu\rho}\,d\,\Omega, \\
 J^{\mu\rho}=\int(\delta x{}^{\mu}\,\tau^{\rho 0}-\delta
x{}^{\rho}\,\tau^{\mu 0}+s^{\mu\rho
0})\,d\,\Omega=\frac{1}{u{}^{0}}(-M^{\mu\rho 0}+M^{\rho\mu
0}+N^{\mu\rho 0}),
\end{array}
\label{I6}
\end{equation}
where  $\tau^{\mu\rho}$ is the energy-momentum tensor for particles,
$s^{\mu\nu\rho}$ is the spin density. The quantity $J^{\mu\rho}$ is
equal to $\int(\delta x{}^{\mu}\,\tau^{kl}-\delta
x{}^{\rho}\,\tau^{\mu\lambda}+s^{\mu\rho\lambda})\,d\,S_{\lambda}$
taken for the volume hypersurface, so it is a tensor, which is
called the {\em total spin tensor}. The quantity $N^{\mu\nu\rho}$ is
also a tensor. The relation $\delta x{}^{0}=0$ gives
$M^{0\nu\rho}=0$. It was assumed that the dimensions of the body are
small, so integrals with two or more factors $\delta x{}^{\mu}$
multiplying $\tau^{\nu\rho}$ and integrals with one or more factors
$\delta x{}^{\mu}$ multiplying $s^{\nu\rho\lambda}$ can be
neglected. The {\em Papapetrou equations of motion for the spin}
(\cite{Pa}-\cite{Mol2}, \cite{Popl}) is
\begin{equation}
\begin{array}{l}
\frac{\stackrel{\circ }{\cal D}}{{\cal
D}s}\,J^{\lambda\nu}=u^{\nu}\,\Theta^{\lambda}-u{}^{\lambda}\,\Theta^{\nu}+
K^{\lambda}_{\phantom{l}\mu\rho}\,N^{\nu\mu\rho}+\frac{1}{2}\,K_{\mu\rho}^{\phantom{ik}\lambda}\,
N^{\mu\nu\rho}-K^{\nu}_{\phantom{j}\mu\rho}\,N^{\lambda\mu\rho}-\frac{1}{2}\,K_{\mu\rho}^{\phantom{ik}\nu}\,N^{\mu\rho\lambda}.
\end{array}
\label{I5}
\end{equation}
Calculating from~(\ref{I8}) the particle 4-acceleration is
\begin{equation}
\begin{array}{l}
\frac{1}{m}\,f^{\mu}_{g}(x)=\frac{d^{2}x{}^{\mu}}{d\widetilde{s}^{2}}=-
\Gamma^{\mu}_{\nu\lambda}\,\left[u{}^{\nu}\,u^{\lambda}+
\frac{1}{u{}^{0}}\,\stackrel{\circ
}{\Gamma}{}^{\phantom{a}\mu}_{\nu\,\rho}\,(u{}^{\nu} \,J^{\rho 0}+
N^{0 \nu\rho})\right]+
\frac{1}{2u{}^{0}}\,K_{\nu\rho}^{\phantom{ik}\mu}\, N^{\nu\rho
0}-\\\frac{1}{2}\,\stackrel{\circ
}{R}{}^{\mu}_{\phantom{j}\nu\sigma\rho}\,u{}^{\nu}\,
J^{\sigma\rho}-\frac{1}{2}\,N_{\nu\rho\lambda}\,K^{\nu\rho\lambda:\,\mu}.
\end{array}
\label{RG1}
\end{equation}
Thus, the relativistic inertial force, exerted on the extended
spinning body moving in the RC space $U_{4}$, can be found to be
\begin{equation}
\begin{array}{l}
\vec{f}_{(in)}(x)=m\vec{a}_{in}(x)=
-\frac{m\vec{a}_{abs}(x)}{\gamma_{q}}=
-m\,\frac{\vec{e}_{f}}{\gamma_{q}}\,\left|\frac{1}{m}\,f^{\alpha}_{(l)}-
\frac{\partial X^{\alpha}}{\partial\, x{}^{\mu}}\,\left[
\stackrel{\circ
}{\Gamma}{}^{\mu}_{\nu\lambda}\,u{}^{\nu}\,u^{\lambda}+
\right.\right.\\\left.\left.  \frac{1}{u{}^{0}}\,\stackrel{\circ
}{\Gamma}{}^{\phantom{a}\mu}_{\nu\,\rho}\,(u{}^{\nu} \,J^{\rho
0}+N^{0 \nu\rho})-
\frac{1}{2u{}^{0}}\,K_{\nu\rho}^{\phantom{ik}\mu}\, N^{\nu\rho
0}+\frac{1}{2}\,\stackrel{\circ
}{R}{}^{\mu}_{\phantom{j}\nu\sigma\rho}\,u{}^{\nu}\,
J^{\sigma\rho}+\frac{1}{2}\,N_{\nu\rho\lambda}\,K^{\nu\rho\lambda:\,\mu}
 \right]\right|.
\end{array}
\label{RLLL10}
\end{equation}
In particular, if the spin density vanishes, $s^{\mu\nu\rho}=0$,
from the conservation law we get then
$\tau^{\mu\rho}=\tau^{\rho\mu}$, $ M^{\mu\rho}=M^{\rho\mu}$,
$M^{\mu\nu\rho}=M^{\mu\rho\rho}$, $N^{\mu\nu\rho}=0$, and that
\begin{equation}
\begin{array}{l}
 J^{\mu\rho}=L^{\mu\rho}=\int(\delta\, x{}^{\mu}\,\tau^{\rho
0}-\delta\, x^{\rho}\,\tau^{\mu
0})\,d\,\Omega=\frac{1}{u{}^{0}}(-M^{\mu\rho 0}+M^{\rho\mu 0}),
\end{array}
\label{I10}
\end{equation}
where $L^{\mu\rho}$ is the angular momentum tensor. The modified
4-momentum (\ref{I7})  reduces to
\begin{equation}
\begin{array}{l}
\Theta^{\nu}=P^{\nu}+\frac{\stackrel{\circ }{\cal D}}{{\cal
D}s}\,L^{\nu\lambda}\,u_{\lambda}.
\end{array}
\label{I11}
\end{equation}
The Eq.~(\ref{I5}) can be recast in the form
\begin{equation}
\begin{array}{l}
\frac{\stackrel{\circ }{\cal D}}{{\cal
D}s}\,L^{\lambda\nu}=u^{\nu}\,\Theta^{\lambda}-u{}^{\lambda}\,\Theta^{\nu},
\end{array}
\label{I12}
\end{equation}
while the  Eq.~(\ref{I8}) becomes
\begin{equation}
\begin{array}{l}
\frac{\stackrel{\circ }{\cal D}\,\Theta^{\nu}}{{\cal
D}\,s}=-\frac{1}{2}\,\stackrel{\circ
}{R}{}^{\nu}_{\phantom{j}\mu\sigma\rho}\,u{}^{\mu}\, L^{\sigma\rho},
\end{array}
\label{I13}
\end{equation}
which give the relativistic inertial force exerted on the spinless
extended body moving in the RC space $U_{4}$ as follows:
\begin{equation}
\begin{array}{l}
\vec{f}_{(in)}(x)=
-m\,\frac{\vec{e}_{f}}{\gamma_{q}}\,\left|\frac{1}{m}\,f^{\alpha}_{(l)}-
\frac{\partial X^{\alpha}}{\partial\, x{}^{\mu}}\,\left[
\stackrel{\circ
}{\Gamma}{}^{\mu}_{\nu\lambda}\,u{}^{\nu}\,u^{\lambda}+
\frac{1}{u{}^{0}}\,\stackrel{\circ
}{\Gamma}{}^{\phantom{a}\mu}_{\nu\,\rho}\,u{}^{\nu} \,L^{\rho
0}+\frac{1}{2}\,\stackrel{\circ
}{R}{}^{\mu}_{\phantom{j}\nu\sigma\rho}\,u{}^{\nu}\, L^{\sigma\rho}
 \right]\right|.
\end{array}
\label{I15}
\end{equation}
If the body is not spatially extended then it is referred to as a
{\em particle}. The corresponding condition $\delta
\,x{}^{\alpha}=0$ gives $ M^{\mu\nu\rho}=0,\phantom{a}
L^{\mu\rho}=0. $ Therefore $\frac{u{}^{\lambda}}{u{}^{0}}\,N^{\mu\nu
0}-N^{\mu\nu\lambda}=0$, which  gives $ N^{\mu\nu\rho}=u{}^{\mu}
\,J^{\nu\rho}, $ so $
J^{\mu\nu}=S^{\mu\nu}=N^{\mu\nu\rho}\,u_{\rho}$, where $S^{\mu\nu}$
is the {\em intrinsic spin tensor}. If the body is spatially
extended then the difference $ R^{\mu\rho}=J^{\mu\rho}-S^{\mu\rho} $
is the {\em rotational spin tensor}. The relativistic inertial force
is then
\begin{equation}
\begin{array}{l}
\vec{f}_{(in)}(x)=
-m\,\frac{\vec{e}_{f}}{\gamma_{q}}\,\left|\frac{1}{m}\,f^{\alpha}_{(l)}-
\frac{\partial X^{\alpha}}{\partial\, x{}^{\mu}}\,\left[
\stackrel{\circ
}{\Gamma}{}^{\mu}_{\nu\lambda}\,u{}^{\nu}\,u^{\lambda}+
\frac{1}{u{}^{0}}\,\stackrel{\circ
}{\Gamma}{}^{\phantom{a}\mu}_{\nu\,\rho}\,(u{}^{\nu} \,S^{\rho 0}+
\overline{u{}^{0}}\,S^{\nu\rho})-
\frac{1}{2u{}^{0}}\,K_{\nu\rho}^{\phantom{ik}\mu}\,
u{}^{\nu}\,S^{\rho 0}+\frac{1}{2}\,\stackrel{\circ
}{R}{}^{\mu}_{\phantom{j}\nu\sigma\rho}\,u{}^{\nu}\,
S^{\sigma\rho}+\right.\right.\\\left.\left.\frac{1}{2}\,u_{\nu}\,S_{\rho\lambda}\,K^{\nu\rho\lambda:\,\mu}
 \right]\right|.
\end{array}
\label{I16}
\end{equation}
In case of the Riemann space, $V_{4} \phantom{a}(\breve{Q}=0)$, the
relativistic inertial force~(\ref{RLLL10}) exerted on the extended
spinning body can be written in terms of the Ricci coefficient of
rotation only:
\begin{equation}
\begin{array}{l}
\breve{\vec{f}}_{(in)}(\breve{x})=
-m\,\frac{\vec{e}_{f}}{\gamma_{q}}\,\left|\frac{1}{m}\,f^{\alpha}_{(l)}-
\frac{\partial X^{\alpha}}{\partial\, \breve{x}{}^{\mu}}\,\left[
\breve{\Gamma}{}^{\mu}_{\nu\lambda}\,\breve{u}{}^{\nu}\,\breve{u}^{\lambda}+
\frac{1}{\breve{u}{}^{0}}\,\breve{\Gamma}{}^{\phantom{a}\mu}_{\nu\,\rho}\,(\breve{u}{}^{\nu}
\,\breve{J}{}^{\rho 0}+  \breve{N}{}^{0 \nu\rho})+
\frac{1}{2}\,\breve{R}{}^{\mu}_{\phantom{j}\nu\sigma\rho}\,\breve{u}{}^{\nu}\,
\breve{J}{}^{\sigma\rho}
 \right]\right|.
\end{array}
\label{I17}
\end{equation}
In case of the Weitzenb\"{o}ck space,
$W_{4}\phantom{a}(\stackrel{\bullet}{R}=0)$, the~(\ref{RLLL10})
reduces to its teleparallel equivalent,
\begin{equation}
\begin{array}{l}
\stackrel{\bullet}{\vec{f}}_{(in)}(\stackrel{\bullet}{x})=
-m\,\frac{\vec{e}_{f}}{\gamma_{q}}\,\left|\frac{1}{m}\,f^{\alpha}_{(l)}-
\frac{\partial X^{\alpha}}{\partial\,
\stackrel{\bullet}{x}{}^{\mu}}\,\left[ \stackrel{\circ
}{\Gamma}{}^{\mu}_{\nu\lambda}\,\stackrel{\bullet}{u}{}^{\nu}\,\stackrel{\bullet}{u}^{\lambda}+
\frac{1}{\stackrel{\bullet}{u}{}^{0}}\,\stackrel{\circ
}{\Gamma}{}^{\phantom{a}\mu}_{\nu\,\rho}\,(\stackrel{\bullet}{u}{}^{\nu}
\,\stackrel{\bullet}{J}{}^{\rho 0}+ \stackrel{\bullet}{N}{}^{0
\nu\rho})-
\frac{1}{2\stackrel{\bullet}{u}{}^{0}}\,\stackrel{\bullet}{K}{}_{\nu\rho}^{\phantom{ik}\mu}\,
\stackrel{\bullet}{N}{}^{\nu\rho
0}+\right.\right.\\\left.\left.\frac{1}{2}\,\stackrel{\bullet}{N}{}_{\nu\rho\lambda}\,\stackrel{\bullet}{K}{}^{\nu\rho\lambda:\,\mu}
 \right]\right|.
\end{array}
\label{I18}
\end{equation}
All magnitudes related with the teleparallel gravity is denoted with
an over '$\bullet$'. Finally, the non-vanishing inertial force,
$\textbf{f}{}_{(in)}^{\,(phot)}(x)$, acting on the photon of energy
$h\nu$ in the $U_{4}$, can be obtained from the ~(\ref{I16}), at
$\vec{f}_{(l)}=0,$ as
\begin{equation}
\begin{array}{l}
\vec{f}{}_{(in)}^{\,(phot)}(x)= -\left(\frac{h\nu}{
c^{2}}\right)\,\vec{e}_{f}\, \left|\frac{\partial
X^{\alpha}}{\partial\,x{}^{\mu}}\,\left[ \stackrel{\circ
}{\Gamma}{}^{\mu}_{\nu\lambda}\,\frac{d\,x{}^{\nu}}{d\,T}\,\frac{d\,x{}^{\lambda}}{d\,T}+
\frac{d\,T}{d\,\overline{t}}\,\stackrel{\circ
}{\Gamma}{}^{\phantom{a}\mu}_{\nu\,\rho}\,(\frac{d\,x{}^{\nu}}{d\,T}
\,S^{\rho 0}+ \frac{d\,\overline{t}}{d\,T}\,S^{\nu\rho})-
\frac{d\,T}{2d\,\overline{t}}\,K_{\nu\rho}^{\phantom{ik}\mu}\,
\frac{d\,x{}^{\nu}}{d\,T}\, S^{\rho
0}+\right.\right.\\\left.\left.\frac{1}{2}\,\stackrel{\circ
}{R}{}^{\mu}_{\phantom{j}\nu\sigma\rho}\,\frac{d\,x{}^{\nu}}{d\,T}\,
S^{\sigma\rho}+\frac{1}{2}\,\frac{d\,x_{\nu}}{d\,T}\,S_{\rho\lambda}\,K^{\nu\rho\lambda:\,\mu}
 \right]\right|,
\end{array}
\label{LL11}
\end{equation}
where $\vec{e}_{f}=(\vec{X}/|\vec{X}|)$, $v_{q}=(\vec{e}_{f}\cdot
\overline{\vec{u}})=|\overline{\vec{u}}|,\,\,(\gamma_{q}=\gamma)$,
$\overline{\vec{u}}$ is the velocity of the photon in $U_{4}$,
 $(d\, \overline{\vec{u}} /d\,\overline{t})$ is the
acceleration, $g_{\mu\nu}\,(d\,x{}^{\mu}/dT)\otimes
(d\,x{}^{\nu}/dT)=0.$

\section{Concluding remarks}
We construct the RTI, which treats the inertia as a distortion of
local internal properties of hypothetical 2D, so-called, {\it
master-space} (MS). The MS is an indispensable companion of
individual particle, without relation to the other matter, embedded
in the background 4D-spacetime. The RTI allows to compute the {\em
inertial force}, acting on an arbitrary point-like observer or
particle due to its {\em absolute acceleration}. In this framework
we essentially improve standard metric and other relevant
geometrical structures related to noninertial frame for an arbitrary
velocities and characteristic acceleration lengths. Despite the
totally different and independent physical sources of gravitation
and inertia, this approach furnishes justification for the
introduction of the principle of equivalence. We relate the inertia
effects to the more general post-Riemannian geometry. We derive a
general expression of the relativistic inertial force exerted on the
extended spinning body moving in the Rieman-Cartan (RC) space.

\end{document}